\documentclass[twocolumn,a4paper,preprintnumbers,floatfix,aps,prb,unsortedaddress,superscriptaddress,longbibliography]{revtex4-2}

\usepackage[para]{threeparttable} 
\usepackage{amsmath}
\usepackage{graphicx} 
\usepackage{booktabs}
\usepackage{subcaption}
\usepackage{url}
\usepackage{hyperref}
\usepackage{miller}
\usepackage{bm}
\usepackage{enumitem}
\usepackage{color}

\usepackage[paperwidth=210mm,paperheight=297mm,centering,hmargin=1.7cm,vmargin=2.5cm]{geometry}

\usepackage[utf8]{inputenc}

\begin{document}

\title{Segregation, ordering, and precipitation in WTaV-based concentrated refractory alloys}

\author{Jesper Byggmästar}
\thanks{Corresponding author}
\email{jesper.byggmastar@helsinki.fi}
\affiliation{Department of Physics, P.O. Box 43, FI-00014 University of Helsinki, Finland}
\author{Damian Sobieraj}
\affiliation{Faculty of Materials Science and Engineering, Warsaw University of Technology, Poland}
\author{Jan S. Wróbel}
\affiliation{Faculty of Materials Science and Engineering, Warsaw University of Technology, Poland}
\author{Daniel K. Schreiber}
\affiliation{Nuclear Sciences Division, Pacific Northwest National Laboratory, United States}
\author{Osman El-Atwani}
\affiliation{Nuclear Sciences Division, Pacific Northwest National Laboratory, United States}
\author{Enrique Martinez}
\affiliation{Departments of Mechanical Engineering and Materials Science and Engineering, Clemson University, United States}
\author{Duc Nguyen-Manh}
\affiliation{Materials Division, United Kingdom Atomic Energy Authority, Culham Campus, Abingdon, United Kingdom}
\affiliation{Department of Materials, University of Oxford, Parks Road, United Kingdom}

\date{\today}

\begin{abstract}
{Tungsten-based low-activation high-entropy alloys are possible candidates for next-generation fusion reactors due to their exceptional tolerance to irradiation, thermal loads, and stress. We develop an accurate and efficient machine-learned interatomic potential for the W--Ta--Cr--V system and use it in hybrid Monte Carlo molecular dynamics simulations of ordering and segregation to all common types of defects in WTaCrV. The predictions are compared to atom probe tomography analysis of segregation and precipitation in WTaCrV thin films. By also considering two other alloys, WTaV and MoNbTaVW, we are able to draw general conclusions about preferred segregation in refractory alloys and the reasons behind it, guiding future alloy design and elucidating experimental observations. We show that the experimentally observed CrV precipitates in WTaCrV form semicoherent bcc-to-bcc interfaces with the surrounding matrix, as coherent precipitates are not thermodynamically stable due to excessive lattice mismatch. The predictions from simulations align well with our atom probe tomography analysis as well as previous experimental observations.}
\end{abstract}

\maketitle

\section{Introduction}
\label{sec:intro}

Concentrated multicomponent alloys, or medium- and high-entropy alloys, of refractory metals have proven to exhibit promising and unique tolerances for conditions of high temperature, stress, or irradiation~\cite{senkov_development_2018,cheng_irradiation_2023}. From the pioneering work of Senkov et al., among the most studied refractory high-entropy alloys are MoNbTaVW and HfNbTaTiZr~\cite{senkov_refractory_2010,senkov_microstructure_2011}. In recent years, the exploration of other alloy compositions has increased. The design of novel alloy compositions may target properties that are crucial for real applications, such as improved ductility or better irradiation tolerance. For example, recent work has revealed several tungsten-based alloys (WTaCrV and WTaCrVHf) as exceptionally irradiation-tolerant materials~\cite{el-atwani_outstanding_2019,el_atwani_quinary_2023}. Recent computational results also revealed V as a critical element controlling radiation damage in refractory alloys and indicate the ternary WTaV alloy to be another, chemically even simpler, radiation-resistant alloy~\cite{wei_revealing_2024}. All these three alloys contain by design no elements prone to high long-term activation due to high-energy irradiation, making them possible candidates for fusion reactors.


Synthesis of high-entropy alloys typically aims for a uniform composition as close as possible to a solid solution, but preferential ordering or segregation may lead to heterogeneous microstructures. This in turn may affect the mechanical properties in unforeseen and nontrivial ways. Short-range order has therefore been studied in refractory alloys in a number of works, using density functional theory (DFT)~\cite{yin_ab_2020}, classical molecular dynamics (MD) and Monte Carlo (MC) simulations~\cite{kostiuchenko_impact_2019-1,li_complex_2020,byggmastar_modeling_2021}, cluster expansion (CE) models~\cite{fernandez-caballero_short-range_2017,sobieraj_chemical_2020,smith_competition_2024} and also experimentally~\cite{fantin_local_2024}. However, segregation to lattice defects is a much less explored topic yet of equal importance. Large-scale atomistic simulations can provide crucial insight, especially with the recent advances on accurate machine-learning (ML) interatomic potentials~\cite{mishin_machinelearning_2021,deringer_gaussian_2021-1,behler_perspective_2016}. Li et al. developed an ML potential for MoNbTaW alloys and studied mechanical loading and grain boundary segregation~\cite{li_complex_2020}. Recently, we developed ML potentials in the tabulated Gaussian approximation potential (tabGAP) framework for MoNbTaVW alloys and used them to study short-range order and segregation in equiatomic MoNbTaVW~\cite{byggmastar_modeling_2021,byggmastar_simple_2022} and high-dose radiation damage in various W-based alloys~\cite{wei_revealing_2024}. Lyu et al.~\cite{lyu_effects_2023} developed an ML potential for WTaCrV and studied how chemical order affects dislocation mobility and other contributors to strengthening, but left segregation unexplored.

In this article we comprehensively study ordering and segregation to defects in the three equiatomic alloys WTaV, WTaCrV, and MoNbTaVW. To this end we first develop an accurate tabGAP machine-learned interatomic potential for the WTaCrV system and validate it for properties relevant for short-range order and segregation. For MoNbTaVW we use a recently developed potential and further validate it. Our aim is to, through systematic simulations and comparison to new and previous experimental analysis, develop a thorough understanding of the segregation and ordering trends in refractory alloys, using three promising W-based alloys as representative test cases. The article begins by the training and validation of the ML potential, followed by the simulations and analysis of short-range order and segregation in the three alloys, and ending with a discussion and comparison to experiments along with the main conclusions.

\section{Methods}
\label{sec:methods}

\subsection{Density functional theory calculations}

DFT calculations for training and testing data are done with the \textsc{vasp} code~\cite{kresse_ab_1993,kresse_ab_1994,kresse_efficiency_1996,kresse_efficient_1996} using projector augmented wave potentials~\cite{blochl_projector_1994,kresse_ultrasoft_1999} (\texttt{\_pv} for Cr and Ta, \texttt{\_sv} for V and W). We use the PBE GGA exchange-correlation functional~\cite{perdew_generalized_1996}, a 500 eV cutoff energy, 0.15 Å$^{-1}$ maximum $k$-point spacing on $\Gamma$-centred Monkhorst-Pack grids~\cite{monkhorst_special_1976}, and 0.1 eV Methfessel-Paxton smearing~\cite{methfessel_high-precision_1989}. The parameters are the same as in our previous work~\cite{byggmastar_modeling_2021}, allowing us to extend the previous training database to include Cr. Spin-polarisation was not considered since WTaCrV alloys do not have a net magnetisation, as concluded previously~\cite{zhao_defect_2020,el-atwani_helium_2020}. Note that pure Cr and dilute Cr alloys are also computed without spin-polarisation, which means that the trained ML potential does not correctly reproduce the Cr ground state~\cite{hafner_magnetic_2002}. The energy and lattice constant of nonmagnetic and antiferromagnetic Cr are very close, but the elastic response differ noticeably~\cite{hafner_magnetic_2002}. For alloys, we investigated the effect of spin-polarisation by relaxing all B2-ordered CrV, CrTa, and CrW binary alloys with initialised antiferromagnetic states at a fixed lattice constant of 3.1 Å. Only CrV shows significant magnetic moments (ferrimagnetic) while CrW is weakly ferrimagnetic. Ferrimagnetic CrV is only 10 meV/atom lower and CrW 0.2 meV/atom lower in energy than in non-spin-polarised relaxations. From this we conclude that spin-polarisation can be neglected when constructing the WTaCrV training database. Further notes on magnetism and energy-volume data are given in Appendix~\ref{sec:app-mag}.

\subsection{WTaCrV interatomic potential}
\label{sec:methods_tabgap}

\subsubsection{Tabulated Gaussian approximation potential}

To develop the WTaCrV ML potential, we use the tabGAP framework~\cite{byggmastar_simple_2022,byggmastar_modeling_2021}. A tabGAP is a Gaussian approximation potential (GAP)~\cite{bartok_gaussian_2010} trained with simple low-dimensional descriptors, which allows the energy predictions to be tabulated onto 1D and 3D grids and evaluated efficiently with cubic spline interpolations~\cite{byggmastar_modeling_2021,glielmo_efficient_2018,vandermause_--fly_2020}. The main benefit of tabGAP is a good balance between computational speed and accuracy, particularly for alloys of many elements~\cite{byggmastar_simple_2022}.

We write the total energy for a system of $N$ atoms as a sum of a pre-fitted fixed pairwise repulsive potential and the ML contributions:
\begin{equation}
    E_\mathrm{tot.} = E_\mathrm{rep.} + E_\mathrm{ML}.
\end{equation}
Hence, for a given DFT training structure, the ML contribution is trained to the energy $E_\mathrm{DFT} - E_\mathrm{rep.}$ (and corresponding forces and virials). The ML potential is a GAP with three different descriptors:
\begin{equation}
    E_\mathrm{ML} = E_\mathrm{2b} + E_\mathrm{3b} + E_\mathrm{EAM}.
    \label{eq:GAP}
\end{equation}
The two-body (2b) descriptor, the interatomic distance $r_{ij}$, produces a ML pair contribution. The three-body (3b) descriptor enumerates all atom triplets $ijk$ using the permutation-invariant vector $[(r_{ij}+r_{ik}), (r_{ij}-r_{ik})^2, r_{jk}]$~\cite{bartok_gaussian_2015}. The 2b and 3b interactions are limited in range with cutoff functions on the $ij$ bonds. The EAM (embedded atom method) descriptor is analogous to the pairwise summed density in EAM potentials~\cite{daw_embedded-atom_1984} and is computed as $\rho_i = \sum_j^N \varphi_{ij} (r_{ij})$, i.e. the descriptor $\rho_i$ is an effective density or coordination of atom $i$. The pair density contribution is here used in the form $\varphi_{ij} (r_{ij}) = (1 - r_{ij} / r_\mathrm{cut})^3$, which smoothly reaches 0 at $r_\mathrm{cut}$. For the simple GAP in Eq.~\ref{eq:GAP}, the three-body term brings most of the accuracy but the EAM term increases stability and improves the accuracy especially for liquids and other different-coordinated structures~\cite{byggmastar_simple_2022}.

GAP uses sparse Gaussian process regression and each descriptor term can be written
\begin{equation}
    E_d = \sum_{i}^{N_d} \delta^2_d \sum_s^{M_d} \alpha_{s} K_d (q_d, q_s),
    \label{eq:GAPd}
\end{equation}
where $d$ represents a descriptor, $N_d$ is the number of descriptor environments (pairs, triplets, or atoms), $\delta^2_d$ is an energy prefactor, and $s$ loops through the $M_d$ number of sparse descriptor environments from the training data. $\alpha_s$ are the regression coefficients to be optimised and $K_d$ is the kernel function that acts as a similarity measure between a known descriptor environment represented by $q_s$ and the desired environment $q_d$. Here we use the squared exponential kernel for all interactions. The total ML energy considers separate terms as Eq.~\ref{eq:GAPd} for each element-specific descriptor, i.e. all unique element pairs for two-body, all element triplets for three-body, and all elements for EAM. 


The repulsive potential is a pre-fitted Ziegler-Biersack-Littmark (ZBL) potential~\cite{ziegler_stopping_1985}
\begin{equation}
    E_\mathrm{rep.} = \sum_{i<j}^N \frac{1}{4\pi \varepsilon_0} \frac{Z_i Z_j e^2}{r_{ij}} \phi_{ij} (r_{ij}/a) f_\mathrm{cut} (r_{ij}),
\end{equation}
where $a = 0.46848 / (Z_i^{0.23} + Z_j^{0.23})$ as in the universal ZBL potential. The screening function for an element pair AB is $\phi_\mathrm{AB}(x) = \sum_{i=1}^m \zeta_i e^{\eta_i x}$. Using $m=3$, the six parameters $\zeta_i$ and $\eta_i$ are fitted to all-electron DFT data from Ref.~\cite{nordlund_repulsive_1997} for each element pair. A cutoff function $f_\mathrm{cut} (r_{ij})$ is used to drive the repulsion to zero before typical bond lengths in near-equilibrium lattices and allow a smooth transition from pair-dominated repulsion to the machine-learned attractive interactions. The latter is ensured by including training data from isolated and embedded dimer repulsion.

After training, the energy predictions from each element pair, triplet, and EAM element are computed and tabulated on 1D and 3D grids and evaluated with cubic spline functions $S$ as
\begin{equation}
\begin{aligned}
     E_\mathrm{tabGAP} =
    & \sum_{i < j}^N S_{\mathrm{rep. + 2b}}^\mathrm{1D} (r_{ij}) + \sum_{i, j<k}^N S_\mathrm{3b}^\mathrm{3D} (r_{ij}, r_{ik}, \cos \theta_{ijk}) \\
    & + \sum_{i}^N S_\mathrm{EAM}^\mathrm{1D} \left(\sum_j^N S_\varphi^\mathrm{1D} (r_{ij}) \right).
\end{aligned}
\end{equation}
Here, the two-body ML contribution and the repulsive potential are merged into one 1D spline over distances (from 0 to $r_\mathrm{cut}$) for each element pair. The three-body cluster term is evaluated with a 3D spline with $(r_{ij}, r_{ik}, \cos \theta_{ijk})$ grid points. The EAM descriptor and energy are tabulated to become conventional EAM potentials. In practice, the pair and EAM terms are tabulated in files compatible with the \texttt{pair\_style eam/fs} in \textsc{lammps} and the three-body grids are stored in a file that requires the \texttt{pair\_style tabgap} available from \url{https://gitlab.com/jezper/tabgap}.


\subsubsection{Training and testing data}

The training data consists of DFT energies, forces, and virials of 13483 structures containing in total 216,118 atoms. The database builds on our previous data for Mo--Nb--Ta--V--W alloys~\cite{byggmastar_modeling_2021,byggmastar_simple_2022}. Even though Cr is the only missing element, instead of extending the database to the six-element system we chose to filter out all Mo- and Nb-containing structures and build a separate WTaCrV database. This is because we aim to make the potential transferable to arbitrary alloy compositions, and sampling a four-element composition space is exponentially easier than a six-element space. The training structures are constructed similarly to the MoNbTaVW dataset (see Supplementary document of \cite{byggmastar_modeling_2021}) by filling in the missing Cr-containing binary, ternary, and quaternary compositions. Even though we consider all alloy compositions, the database is somewhat biased towards quaternary WTaCrV alloys. The training structures include defect-free BCC crystals at different volumes, structures with vacancies, self-interstitial atoms, liquids, a few surface structures, and ordered structures. For more details about the construction of training structures, see Supplementary document of \cite{byggmastar_modeling_2021}. Additionally, a small set of pure Cr structures are included by rescaling a subset of the pure W structures~\cite{byggmastar_machine-learning_2019}. Compared to the MoNbTaVW database~\cite{byggmastar_modeling_2021,byggmastar_simple_2022}, where reduced pure-element databases from Ref.~\cite{byggmastar_gaussian_2020} were used, here we used the full sets of W, Ta, and V structures. This is to ensure that the A15 crystal structure is not incorrectly stabilised over the ground-state BCC phase, an issue recently pointed out~\cite{starikov_angulardependent_2024}.

Since we aim to investigate order and segregation, we also included all BCC-like ordered binaries discovered and considered in Ref.~\cite{sobieraj_chemical_2020}. In addition, we included intermetallic Laves (C14, C15, C36) and A15 phases of all binary alloys to reproduce the correct thermodynamic ground states for V-Ta and Cr-Ta. Like in Ref.~\cite{byggmastar_modeling_2021}, we also performed iterative training to ensure good accuracy for short-range order in the quaternary alloys. This was done by using a preliminary version of the potential in MC/MD simulations to produce defect-free WTaCrV with short-range order, as well as local segregation in systems with vacancies or self-interstitial atoms.

The hyperparameters for the training were identical to our previous work on the MoNbTaVW system~\cite{byggmastar_simple_2022} and input files can be found in Ref.~\cite{byggmastar_2025_zenodo}. In short, the cutoff distance for the two-body and EAM terms is 5 Å and for the three-body term 4.1 Å. The number of sparse points, $M_d$, are 20 for two-body and EAM descriptors and 300 for three-body descriptors, picked using uniform sampling. The regularisation parameters $\sigma$ in GAP are set lowest for BCC crystals (2 or 5 meV/atom for energy, 0.1 eV/Å for force, and 0.2 or 0.5 eV for virials) to achieve best accuracy for them. For liquids, surfaces, intermetallic and other non-BCC structures considered less important, the regularisation parameters are higher. For the full input, see Ref.~\cite{byggmastar_2025_zenodo}. The tabulation of the GAP energy predictions to the tabGAP uses 5000 points for the 1D grids (pair potential and EAM functions) and an $80\times80\times80$ grid for the three-body $(r_{ij}, r_{ik}, \cos \theta_{ijk})$ grid, which results in spline interpolation errors that are negligible in comparison to the training errors~\cite{byggmastar_modeling_2021}.

\subsection{Formation and migration energy calculations}

For calculations of formation energies of bulk random alloys and point defects in WTaCrV, the systems were first fully relaxed to zero pressure and minimum energy with the conjugate-gradient method in \textsc{lammps} or \textsc{vasp}. The force tolerance for relaxation was 0.01 eV/Å in both cases for consistency. The formation energy of bulk alloys is then obtained as
\begin{equation}
    E_\mathrm{f} = E_\mathrm{alloy} - \sum_A N_A E_A,
    \label{eq:ef}
\end{equation}
where $E_\mathrm{alloy}$ is the total energy of the relaxed alloy, $N_A$ is the number of atoms of element $A$ in the alloy, and $E_A$ is the total energy per atom of $A$ in its reference ground state (pure BCC for all elements W, Ta, Cr, V). 

The formation energies of single vacancies and self-interstitials in WTaCrV is calculated with the tabGAP as
\begin{equation}
    E_\mathrm{f} = E_\mathrm{def} - E_\mathrm{bulk} \pm \mu_A,
\end{equation}
with $+\mu_A$ for vacancies and $-\mu_A$ for interstitials. $E_\mathrm{def}$ is the total energy of the relaxed system with the defect, $E_\mathrm{bulk}$ is the corresponding defect-free alloy and $\mu_A$ is the chemical potential of the element $A$ (the removed atom for vacancy or one of the dumbbell atoms for self-interstitials). The system size is $4\times 4\times 4$ cubic BCC unit cells. 

The chemical potentials for each element in equiatomic WTaCrV were determined by a substitution method similar to Refs.~\cite{zhao_defect_2020,jeffries_prediction_2025}, where in equiatomic alloy systems one atom is replaced by each of the remaining elements. After relaxation this provides four energies with slightly varying chemical compositions. We performed substitutions and relaxations in 1000 different random alloy systems, yielding 4000 total energies collected in a vector $\mathbf{E}$. The chemical potentials $\bm{\mu}$ are extracted from the least-squares linear fit $\mathbf{E} = \mathbf{N} \bm{\mu}$, where $\mathbf{N}$ is the matrix of compositions (number of atoms of each element) in the relaxed systems. The results are $\mu_\mathrm{Cr} = -9.388$ eV, $\mu_\mathrm{Ta}=-11.724$ eV, $\mu_\mathrm{V}=-9.033$ eV, and $\mu_\mathrm{W}=-13.029$ eV. Tests showed that these values are converged to within 0.02 eV with respect to the sampling statistics. The chemical potentials obtained for the tabGAP here are very close to the DFT values in~\cite{zhao_defect_2020}, indicating that both the tabGAP and the chemical potential fitting method are reliable. The chemical potentials are close to the pure-metal BCC energies ($-9.515$, $-11.818$, $-8.995$, and $-12.955$ eV for Cr, Ta, V, W), which are often used as the chemical potentials and would here introduce errors on the order of 0.1 eV only. We note that we here only consider random alloys, corresponding to the high-temperature limit where short-range order vanishes and the results are directly comparable to available DFT data for vacancy formation energies~\cite{zhao_defect_2020}. At finite temperature $T$, the free energy of vacancy formation is traditionally
estimated by resorting to Widom’s technique. This approach involves sampling over the alloy configurations and exponential averaging of the difference of vibrational free
energy associated with the creation of a vacant site via the transfer of an A atom to an external reservoir whose chemical potential is $\mu_{A}$ \cite{wrobel_2025}. The Widom technique accounting for short-range order when computing chemical potentials and free energy of vacancy formation energies has recently also been discussed for binary alloys ~\cite{schuler_accurate_2024,li_vacancy_2024}.

The calculation of formation energies for self-interstitials in random alloys is not trivial. Regardless of how the inititial structure is constructed, during relaxation the inserted interstitial often relaxes far from the initial position before finding a local minimum~\cite{zhao_defect_2020,byggmastar_modeling_2021}. This means that the interstitial-free bulk in which the interstitial is inserted is no longer the proper reference structure and the inserted element is not necessarily part of the relaxed dumbbell configuration. Here we therefore construct the reference defect-free system \textit{after} relaxation of the dumbbell. This is done by first identifying the dumbbell atoms with Wigner-Seitz analysis in \textsc{ovito}~\cite{stukowski_visualization_2010}, then choosing one dumbbell atom as part of the reference bulk and one as the interstitial. Given a dumbbell, we choose the interstitial element with priority in the order of atom size: Cr, V, W, Ta. To create the reference bulk, the interstitial atom is removed and the other atom is inserted at the midpoint of the dumbbell bond. After relaxation this is a defect-free BCC lattice and a proper reference for the formation energy calculation. For both vacancies and interstitials, we relaxed 10 000 different WTaCrV systems to obtain distributions of the formation energies.

The migration energies of single vacancies were computed with the climbing-image NEB method~\cite{henkelman_climbing_2000} as implemented in \textsc{lammps}. The system size was 128 atoms. In total, 22,500 migration energies for first-nearest neighbour jumps were obtained with each barrier calculated in a different randomly ordered WTaCrV lattice.

Surface and grain boundary energies are calculated by relaxing the corresponding structure and computing the energy penalty per area (surface or grain boundary) compared to pristine bulk. The different grain boundary structures were obtained from~\cite{zheng_grain_2020} and relaxed with the tabGAPs.

\subsection{Hybrid MC/MD simulations}

Hybrid MD/MC simulations are done with the \textsc{lammps} code~\cite{thompson_lammps_2022}. The MD integration is always carried out in the $NPT$ ensemble to allow the pressure to relax to zero, except for the surface structures which were in $NVT$. Simulations are done to investigate short-range order in single crystals and segregation to various lattice defects: grain boundaries in polycrystals, surfaces, voids, dislocation lines, dislocation loops, and precipitates. In all cases we consider three equiatomic alloys WTaCrV, WTaV, and MoNbTaVW, initially randomly ordered. For all simulations of WTaCrV and WTaV we use the tabGAP developed here and for MoNbTaVW the tabGAP from~\cite{byggmastar_simple_2022}. Note that WTaV could be simulated with both potentials. For a few cases we tested that the predicted order and segregation trends are the same in both potentials. In Appendix~\ref{sec:app-cross} we additionally show cross-validation of the two tabGAPs for WTaV between each other and DFT, showing that the differences between the three are only a few meV/atom, as expected from the train and test accuracy. Polycrystalline structures were constructed using ATOMSK~\cite{hirel_atomsk_2015}, surfaces using ASE~\cite{larsen_atomic_2017}, and all other structures using self-made scripts.

In all MC/MD simulations, MC swap attempts are performed in the canonical ensemble for 1\% of the atoms every 10 MD steps. The MC/MD simulations for single crystals are continued until the potential energy does not significantly decrease. This required between 4 and 8 million MD steps for low temperatures (100--300 K) and significantly fewer for higher temperatures, although at least 1 million MD steps was done for all cases. For defect segregation simulations, all simulations are done to 100,000 MD steps at 300 K. This proved to be enough for segregation around the defects to emerge and stabilise. Even though the systems do not reach equilibrium and the chemical order in bulk regions far from the defect are not fully optimised, in this case our focus is solely on the segregation close to the defects. Visualisations and analysis are done with \textsc{ovito}~\cite{stukowski_visualization_2010}.

After the MC/MD relaxations of the single crystals, the short-range order (SRO) parameters for each pair $AB$ were computed as
\begin{equation}
 S_{\Delta r_{ij}}^{AB} = 1 - \frac{p_{\Delta r_{ij}}^{AB}}{c_B},
\end{equation}
where $\Delta r_{ij}$ is the distance range of the neighbour shell to consider, $p_{\Delta r_{ij}}^{AB}$ is the conditional probability of finding the element $B$ in the neighbour shell of $A$, and $c_B$ is the concentration of $B$ in the alloy. Here we consider first-nearest neighbour pairs using $\Delta r_{ij} = [0, 3]$ Å for WTaV and MoNbTaVW and $[0, 2.9]$ Å for WTaCrV, and the second-nearest shell using $\Delta r_{ij} = [3, 3.9]$ Å for all three alloys. Negative $S_{\Delta r_{ij}}^{AB}$ indicates attraction (i.e. more $AB$ pairs than expected in a random mixture) and positive values repulsion (fewer $AB$ pairs).

\subsection{Atom probe tomography analysis}

For the APT analysis we revisited the W$_{38}$Ta$_{36}$Cr$_{15}$V$_{11}$ data from the thin film samples from Ref.~\cite{el-atwani_outstanding_2019}, both before and after irradiation. The sample preparation and irradiation conditions are described in detail in Ref.~\cite{el-atwani_outstanding_2019}. In short, magnetron sputtering was used to produce the films. Irradiation was performed with 3 MeV Cu$^+$ ions to a dose of roughly 8 dpa. Samples for APT were prepared using established focused ion beam liftout and annular milling methods~\cite{thompson_situ_2007}. APT analyses were performed with a CAMECA LEAP 4000X HR in laser pulsing mode ($\lambda$ = 355 nm, 40 pJ/pulse). The samples were held at 45 K in ultra-high vacuum ($1.1\times 10^{-11}$ Torr) during data collection. APT data were reconstructed using the CAMECA IVAS software, version 3.8.14. Isoconcentration surfaces and associated proximity histograms were generated with an isotropic 3D grid consisting of 0.8 nm voxels and 2.4 nm delocalization~\cite{hellman_analysis_2000}.

\subsection{Interface and coherency strain energies}

To understand the APT results, we computed the interface energetics between the most relevant B2-ordered binary alloys of the W--Ta--Cr--V system. The formation energy per atom of a periodic interfacial superlattice of two slabs $A$ and $B$ is~\cite{ozolins_effects_1998,wolverton_shortrangeorder_2000}
\begin{equation}
   \frac{E_\mathrm{f}}{N} = \frac{2S\sigma}{N} + \zeta + \frac{\Delta E_\mathrm{f}^{AB}}{N},
   \label{eq:Ef_int}
\end{equation}
where the first term is the energy for creating the interface with $S$ as the interface area ($2S$ due to periodic boundaries) and $\sigma$ as the interfacial energy per area unit. The second term, $\zeta$, is the energy per atom due to straining the two lattices $A$ and $B$ to a given (or relaxed) lattice size. The third term is the formation energy of the relaxed and separated $A$ and $B$ lattices with respect to relaxed reference ground states. If $A$ and $B$ are pure metals, or if the formation energy is computed with $A$ and $B$ as the references, the third term is zero. In our case, $A$ and $B$ are binary B2-ordered lattices and the references are the relaxed pure BCC metals. By creating and relaxing interfacial superlattices and computing the formation energy from Eq.~\ref{eq:ef}, Eq.~\ref{eq:Ef_int} can be used to solve for either the interfacial energy $\sigma$ or the strain energy $\zeta$. 

If $A$ and $B$ are joined to form a coherent interface, both lattices must be strained to shared lattice spacings perpendicular to the interface normal. This coherency strain energy is given by~\cite{wolverton_shortrangeorder_2000}
\begin{equation}
    \Delta E_\mathrm{CS}(\hat{k}, x) = \min_{a_\perp} \left[ (1-x)\Delta E_A(\hat{k}, a_\perp) + x \Delta E_B(\hat{k}, a_\perp)  \right],
    \label{eq:Ecoh}
\end{equation}
where $\hat{k}$ is the interface normal direction, $x$ is the composition (here $x=0.5$). $\Delta E_A(\hat{k}, a_\perp)$ and $\Delta E_B(\hat{k}, a_\perp)$ are the energies required to strain $A$ and $B$ from equilibrium lattice constants to $a_\perp$ perpendicular to the interface normal while allowing for relaxation parallel to the interface normal. The energy is minimised with respect to $a_\perp$ to find the optimal coherent interface. 

We determine the coherency strain energy and $a_\perp$ with both DFT and the tabGAP for CrV/TaW and CrTa/VW interfaces. All binary alloys are in B2 order. We consider $\hkl(100)_A||\hkl(100)_B$, $\hkl(110)_A||\hkl(110)_B$, and $\hkl(111)_A||\hkl(111)_B$ interfaces, where $A$ and $B$ denote the binary alloy pairs. With tabGAP, we sample 100 trial $a_\perp$ between 2.8 and 3.2 Å for $A$ and $B$, relax each lattice along the interface direction (separately for each interface), and use Eq.~\ref{eq:Ecoh} to find the coherency strain energy and optimal $a_\perp$ for $x=0.5$. In DFT, we use tabGAP-relaxed lattices of $A$ and $B$ for 20 trial $a_\perp$ in the same range as input. For each $a_\perp$ lattice, we strain the cell slightly in the interface directions until the energy minimum is found. After that, $\Delta E_\mathrm{CS}$ and the optimal $a_\perp$ are obtained.

We also compute the coherency strain energy for 3D precipitates of CrV or CrTa in TaW or VW matrices using the same methods, but with isotropic strain to match the two lattices coherently in all directions. For this, sampling the bulk energy-volume equation of states for the B2 binaries is enough to obtain the 3D coherency strain energy $\Delta E_\mathrm{CS}^\mathrm{3D}$.

For coherent interfaces of given geometry, the strain energy $\zeta = \Delta E_\mathrm{CS}$ in Eq.~\ref{eq:Ef_int}. The interfacial energy $\sigma$ can then be solved from Eq.~\ref{eq:Ef_int} using the formation energy of a relaxed interfacial superlattice. With the tabGAP, we create superlattices of all the above B2 interfaces with increasing length of the periodic cell, relax the cells and atom positions, and calculate $\sigma$. We found that $\sigma$ converges rapidly with respect to the superlattice length, so that in DFT we could use small supercells of 20--25 Å in length along the interface normal for the \hkl(100), \hkl(110), and \hkl(111) interfaces. The cells are fully relaxed in DFT and the interfacial energies computed in the same way as with tabGAP.

For CrV/TaW, the lattice mismatch is large and semicoherent interfaces with misfit dislocations in the interface can be expected. The lattice mismatch is minimised when joining 10 unit cells of CrV with 9 unit cells of TaW. For semicoherent interfaces, we assume that each lattice can fully relax so that $\zeta = 0$. We then obtain the interfacial energy associated with the interface and misfit dislocations from Eq.~\ref{eq:Ef_int} as before, after full relaxation of the interface system. Here, the system size is too large for DFT so only tabGAP is used. We also created larger semicoherent interfacial cells (20 CrV and 18 TaW unit cells) for MC/MD relaxations with the tabGAP.

\section{Results}
\label{sec:results}

\begin{figure*}
    \centering
    \includegraphics[width=\linewidth]{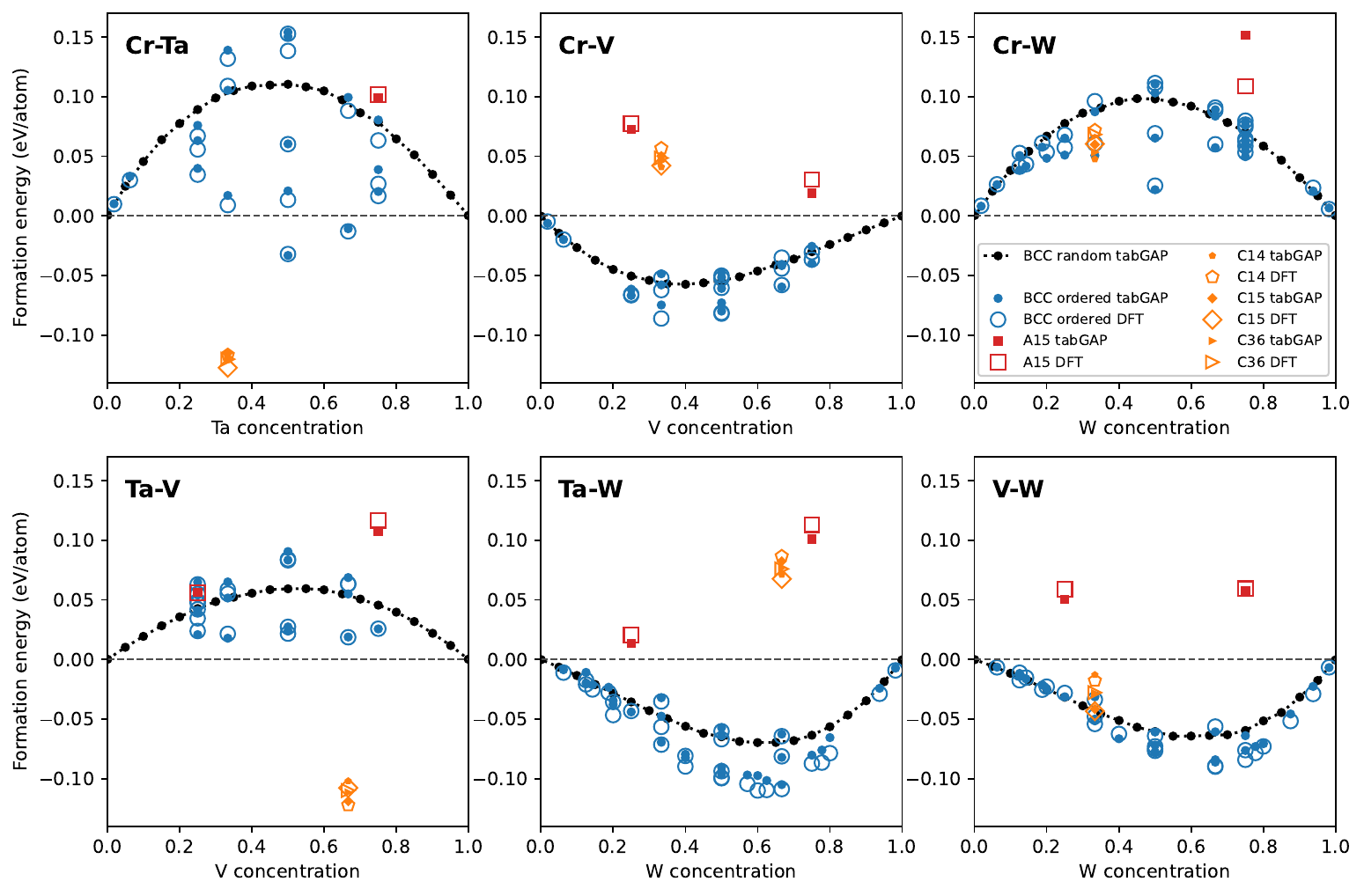}
    \caption{Formation energies of relaxed binary alloys from DFT compared with the W--Ta--Cr--V tabGAP.}
    \label{fig:binaries}
\end{figure*}

\subsection{Machine-learned interatomic potential}

The train and test errors for arbitrary crystalline alloy compositions are a few meV/atom for energies and around 0.1 eV/Å for force components. However, even such low values can lead to significant errors in physically important properties, such as formation energies of bulk alloys, point defects, surfaces, or grain boundaries. We here first validate the tabGAP for the properties most relevant for ordering and segregation simulations. The properties calculated here will thereafter be useful when interpreting the results of the short-range order, segregation, and precipitation simulations.

\subsubsection{Binary alloys}

Figure~\ref{fig:binaries} shows formation energies of binary alloys, including BCC-like random structures and ordered phases (from Ref.~\cite{sobieraj_chemical_2020}) as well as the A15 and Laves C14, C15, and C36 intermetallic phases. The formation energies of random alloys are the averages of three different 1024-atom systems. All structures are fully relaxed (positions and cell shape) in both DFT and with the tabGAP. Note that similar structures are also included in the training database, however they are not relaxed. Figure~\ref{fig:binaries} shows that the tabGAP reproduces DFT formation energies of the BCC ordered and intermetallic structures overall well. We found it important to ensure that the experimentally known Laves phases of V$_2$Ta and Cr$_2$Ta are correctly predicted as the lowest-energy phase, as BCC-like structures for these binaries are mostly thermodynamically unstable. Even though the tabGAP does predict accurate formation energies for the Laves phases, the order of the C14, C15, and C36 phases are not always in agreement with DFT. The Laves phases for other binaries are not thermodynamically stable, and the A15 phase is never stable. From Fig.~\ref{fig:binaries} we also note that the difference between random BCC alloys and the lowest-energy ordered BCC phases from Ref.~\cite{sobieraj_chemical_2020} are often significant, indicating strong tendencies for short-range order.

\begin{figure}
    \centering
    \includegraphics[width=\linewidth]{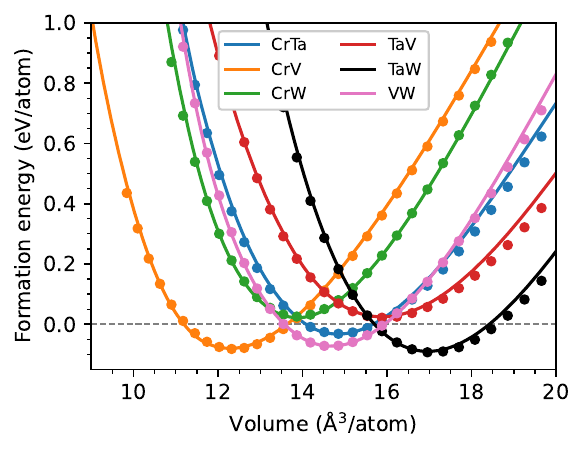}
    \caption{Energy as a function of atomic volume for binary B2-ordered alloys. Lines are tabGAP predictions and points are DFT.}
    \label{fig:B2}
\end{figure}

The training database only includes one structure for every ordered binary composition with a volume close to but not relaxed to zero pressure. To confirm that the tabGAP can predict the correct elastic response, Fig.~\ref{fig:B2} shows energy-volume curves for all binary alloys in the B2 (CsCl) ordered phase. Even though the B2 phase is not necessarily the ground state of all binaries, it is generally relatively low in energy and therefore good representative test systems. Figure~\ref{fig:B2} shows that the agreement between tabGAP and DFT is good for all binaries, with visible deviations only at large atomic volumes.

\subsubsection{WTaCrV alloys}

\begin{figure}
    \centering
    \includegraphics[width=\linewidth]{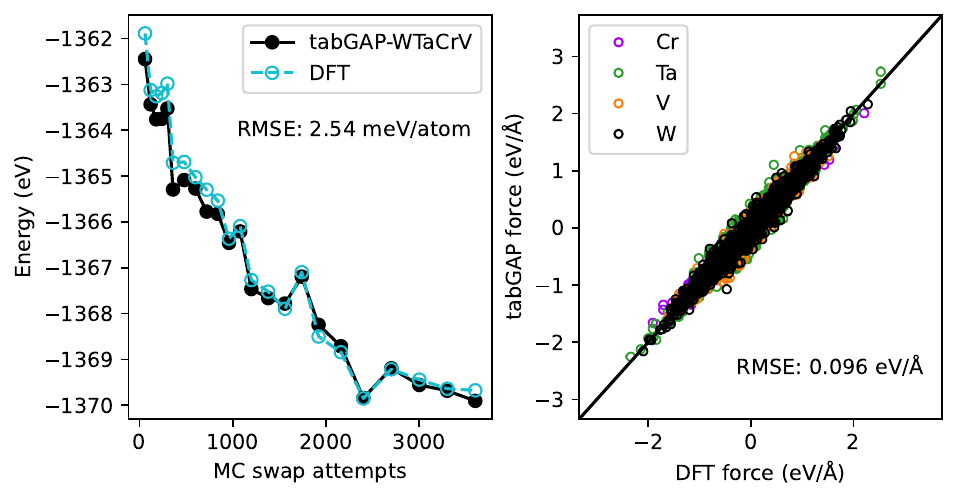}
    \caption{Energy evolution during a MC/MD simulation of a 127-atom WTaCrV system with one vacancy with the tabGAP (left). DFT energies are obtained for comparison. The energy and (right) force RMSEs between tabGAP and DFT are calculated by combining all frames during the simulation.}
    \label{fig:vasp_WTaCrV}
\end{figure}

We used a hybrid MC/MD simulation as a test case for how accurate the tabGAP is for short-range ordered structures of WTaCrV. Figure~\ref{fig:vasp_WTaCrV} shows the energy as a function of number of MC swap attempts for a tabGAP MC/MD simulation at 300 K of a 128-atom initially random WTaCrV system. The sample contains one vacancy to trigger some local segregation in addition to bulk short-range order. At regular intervals, we extract a structure from the simulation and calculate the DFT energy. Figure~\ref{fig:vasp_WTaCrV} shows that the energies of tabGAP and DFT closely match, with an RMSE of 2.54 meV/atom across all frames. Figure~\ref{fig:vasp_WTaCrV} also shows a parity plot of tabGAP and DFT force components for all frames. With force components ranging from $-2$ to 2 eV/Å, the RMSE is lower than 0.1 eV/Å.

\begin{figure}
    \centering
    \includegraphics[width=\linewidth]{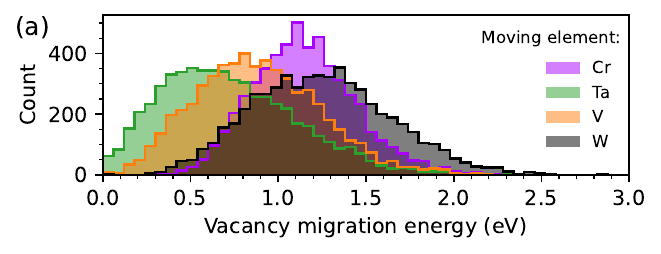}
    \includegraphics[width=\linewidth]{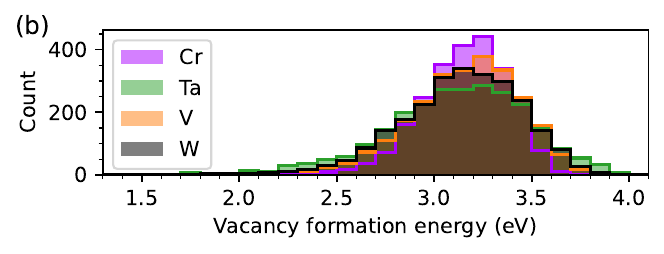}
    \includegraphics[width=\linewidth]{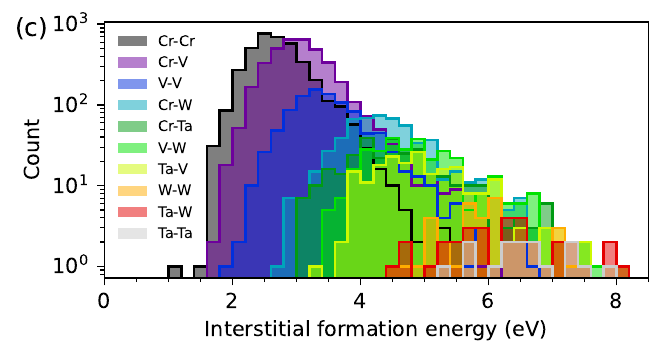}
    \caption{Distributions of (a) single-vacancy migration energies, (b) formation energies, and (c) stable self-interstitial dumbbell formation energies in equiatomic WTaCrV calculated with the tabGAP. The distributions are plotted separately for each element, which for migration is the element that exchanges position with the vacancy, for vacancy formation the element that is removed, and for self-interstitials the two elements of the dumbbell configuration. The legend for interstitial dumbbells is given in order of the average formation energy (Cr--Cr lowest, Ta--Ta highest). Note the logarithmic scale for interstitials, since the vast majority of stable dumbbells are Cr-- or V--containing pairs.}
    \label{fig:vacmig}
\end{figure}

\begin{table}
    \centering
    \caption{Average vacancy migration energies ($E_\mathrm{m}^\mathrm{vac}$, eV) as well as average vacancy and self-interstitial formation energy ($E_\mathrm{f}$, eV) in equiatomic randomly ordered WTaCrV. The migration energies are separated by the element that exchanges position with the first-nearest neighbour vacancy.}
    \begin{tabular}{llll}
     \toprule
      & Element & DFT~\cite{zhao_defect_2020} & tabGAP \\
     \midrule
     $E_\mathrm{m}^\mathrm{vac}$ & W & 1.15 & $1.25 \pm 0.40$  \\
     $E_\mathrm{m}^\mathrm{vac}$ & Ta & 0.71 & $0.72 \pm 0.39$ \\
     $E_\mathrm{m}^\mathrm{vac}$ & Cr & 1.11 & $1.15 \pm 0.30$ \\
     $E_\mathrm{m}^\mathrm{vac}$ & V & 1.00 & $0.89 \pm 0.36$ \\
     $E_\mathrm{m}^\mathrm{vac}$ & all & - & $1.00 \pm 0.42$ \\
     $E_\mathrm{f}^\mathrm{vac}$ & all & 3.18 & $3.14 \pm 0.30$ \\ 
     $E_\mathrm{f}^\mathrm{SIA}$ & all & - & $3.25 \pm 0.91$ \\
     \bottomrule
    \end{tabular}
    \label{tab:vacmig}
\end{table}

For energetically favourable short-range order or segregation to develop, atoms must move from their lattice sites. Since in MC/MD simulations the kinetics and time scale that correspond to the emerging order or segregation are not accessible, it is important to also consider the mobility of atoms. Hence we calculate migration energies of single vacancies in equiatomic WTaCrV with the tabGAP, for which DFT data are available for comparison and validation~\cite{zhao_defect_2020}. We also calculated formation energies of single vacancies and self-interstitial atoms. Figure~\ref{fig:vacmig} shows distributions of the results. As observed previously~\cite{zhao_defect_2020}, the migration and formation energies fall on wide distributions. The vacancy migration energies depend strongly on the element that exchanges position with the vacancy as well as the local chemical environment. The formation energies are also strongly dependent on the latter.

Self-interstitials are on average most stable as Cr--Cr, Cr--V, and V--V dumbbells in that order. Ta- and W-containing dumbbells are rarely stable. From relaxations of 10,000 interstitial systems, only 15 cases relaxed to Ta--Ta dumbbells, 34 to Ta--W, and 41 to W--W. For comparison, the number of Cr--Cr, Cr--V, and V--V dumbbells were 3825, 3720, and 958, respectively. The most common dumbbell directions are \hkl<110> and directions around \hkl<221> and \hkl<111>, again strongly dependent on the local chemical environment. All these results are consistent with the DFT observations~\cite{zhao_defect_2020}. Quantitative comparison between DFT and the tabGAP for average migration and formation energies is shown in Tab.~\ref{tab:vacmig}. The agreement between tabGAP and DFT is excellent, especially considering the limited statistics of the DFT data and the wide distributions.

\subsubsection{Pure-element properties}

\begin{table}
    \centering
    \caption{Lattice constants (Å), surface energies (meV/Å$^2$), and grain boundary energies (meV/Å$^2$) of the pure BCC elements as predicted by the tabGAPs and compared with DFT results (in italic). The DFT lattice constants are calculated here and the surface and grain boundary energies are from Ref.~\cite{tran_surface_2016,zheng_grain_2020}.}
    \begin{tabular}{lcccccc}
     \toprule
      & W & Ta & Cr & V & Mo & Nb \\
     \midrule
     $a$ & 3.184 & 3.324 & 2.843 & 3.000 & 3.158 & 3.309 \\
         & \textit{3.185} & \textit{3.319} & \textit{2.846} & \textit{2.997} & \textit{3.163} & \textit{3.307}  \\
     $\gamma_\mathrm{\hkl(110)}$ & 201 & 148 & 200 & 148 & 175 & 140 \\
     & \textit{202} & \textit{146} & \textit{201} & \textit{151} & \textit{174} & \textit{129} \\
     $\gamma_\mathrm{\hkl(100)}$ & 233 & 170 & 215 & 164 & 193 & 163 \\
     & \textit{247} & \textit{154} & \textit{227} & \textit{149} & \textit{199} & \textit{142} \\
     $\gamma_\mathrm{\hkl(111)}$ & 237 & 181 & 237 & 177 & 206 & 167 \\
     & \textit{216} & \textit{169} & \textit{214} & \textit{169} & \textit{185} & \textit{146} \\
     $\gamma_{\Sigma\mathrm{3}(110)}$ & 58 & 28 & 53 & 23 & 44 & 25 \\
     & \textit{45} & \textit{21} & \textit{42} & \textit{20} & \textit{31} & \textit{18} \\ 
     $\gamma_{\Sigma\mathrm{3}(112)}$ & 50 & 24 & 46 & 21 & 39 & 21 \\
     & \textit{42} & \textit{18} & \textit{40} & \textit{16} & \textit{30} & \textit{16} \\
     \bottomrule
    \end{tabular}
    \label{tab:props}
\end{table}

\begin{figure}
    \centering
    \includegraphics[width=\linewidth]{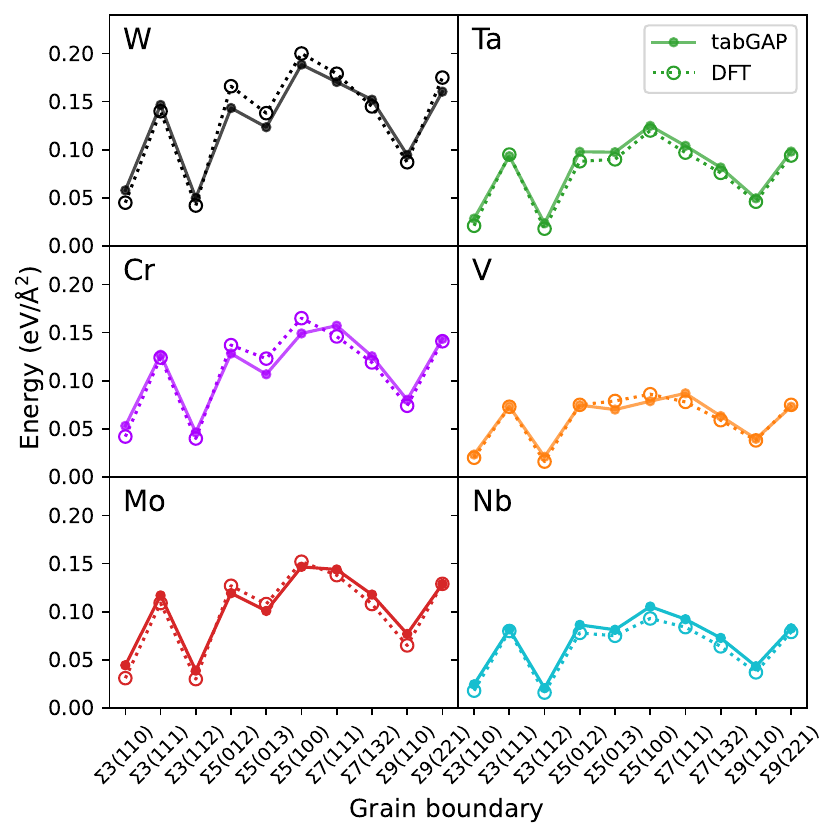}
    \caption{Relaxed grain boundary energies of all six pure metals compared between DFT and the tabGAPs.}
    \label{fig:gbs}
\end{figure}

Even though the primary purpose of the WTaCrV tabGAP is to simulate alloys, it can also be used in simulations of the pure metals. More importantly, some fundamental pure-element properties can be used to either predict or be directly linked to the observed segregation behaviour. Table~\ref{tab:props} summarises some of the key pure-metal properties for the later segregation studies; lattice constant of BCC, surface energies, and grain boundary energies. The Mo and Nb properties are calculated with the MoNbTaVW tabGAP and all other elements with the new WTaCrV tabGAP. Overall, the agreement between tabGAP and DFT is favourable. The noteworthy overall trends across elements are: (1) the order of the lattice constants is Cr $<$ V $<$ Mo $\simeq$ W $<$ Nb $\simeq$ Ta, (2) the order of surface energies is Nb $<$ V $\simeq$ Ta $<$ Mo $<$ Cr $\simeq$ W, and (3) the order of grain boundary energies is Nb $\simeq$ V $<$ Ta $<$ Mo $<$ Cr $<$ W. The order of surface energies is not universal for all elements in DFT. \hkl(110) is lowest in energy for all elements except V. For group 6 elements (Cr, Mo, W), \hkl(111) is lower in energy than \hkl(100) and vice versa for group 5 elements (V, Nb, Ta). This trend is not captured by the tabGAPs, which favours \hkl(110) over \hkl(100) over \hkl(111) in all elements. 

Figure~\ref{fig:gbs} shows relaxed interface energies of 10 grain boundaries. Figure~\ref{fig:gbs} shows that the tabGAPs reproduce the DFT grain boundary energies well. Most importantly, the $\Sigma\mathrm{3}(112)$ is lowest in energy closely followed by $\Sigma\mathrm{3}(110)$ for all elements in both DFT and the tabGAPs. Note that no grain boundary structures were included in the training data.

\subsection{Short-range order}
\label{sec:sro}

\begin{figure*}
    \centering
    \includegraphics[width=\linewidth]{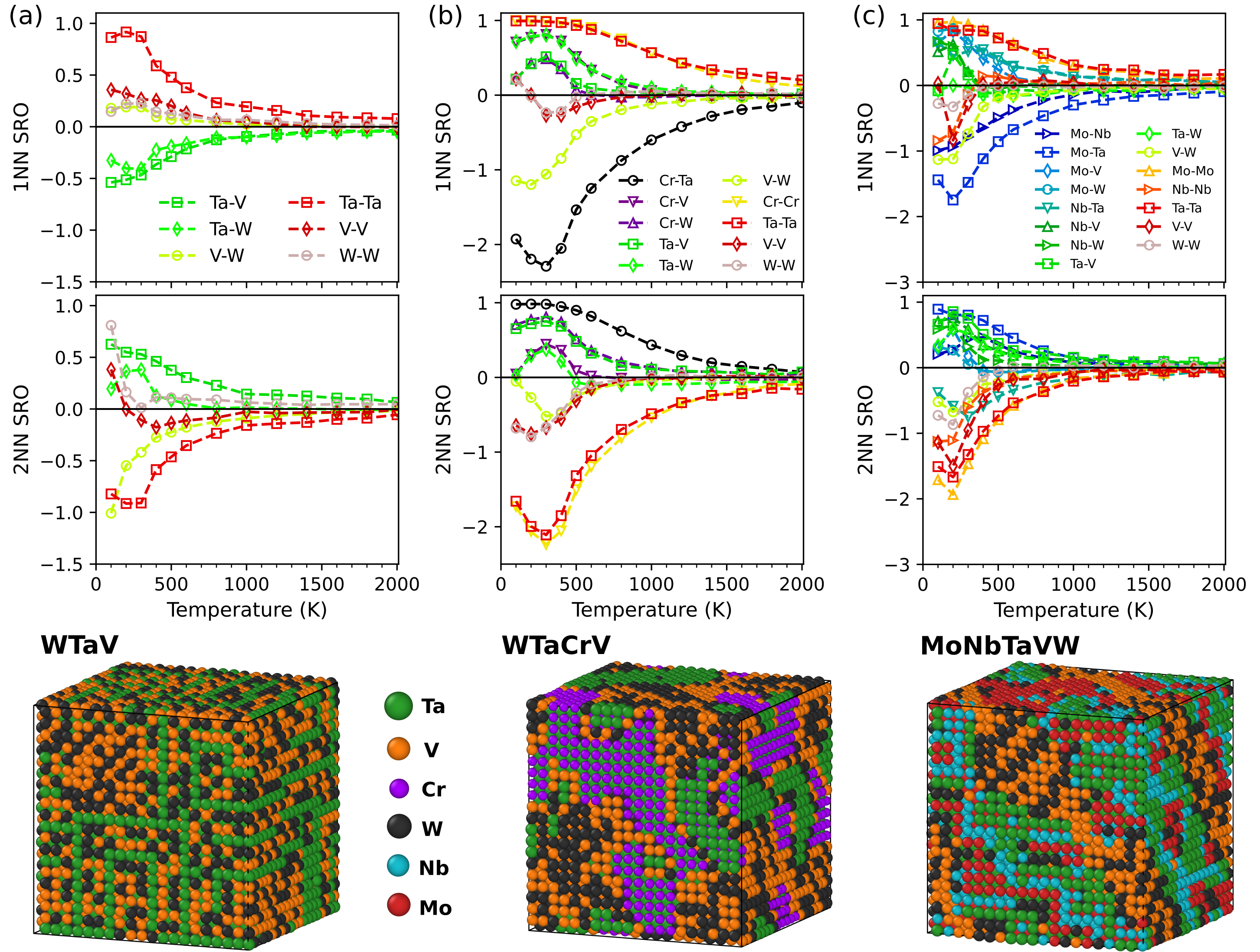}
    \caption{Short-range order parameters for first and second-nearest neighbour shells, 1NN and 2NN, as functions of temperature in single-crystal MC/MD-relaxed BCC systems (16 000 atoms) of (a) WTaV, (b) WTaCrV, and (c) MoNbTaVW. The snapshots show the final structures at 200 K.}
    \label{fig:sro}
\end{figure*}

Having validated that the tabGAPs perform well for properties relevant to order and segregation, we begin by analysing the predicted short-range order in single-crystal alloys of WTaV, WTaCrV, and MoNbTaVW. The results from MC/MD relaxations are summarised in Fig.~\ref{fig:sro}, showing SRO parameters for the first- and second-nearest neighbour shells and snapshots of the MC/MD-relaxed structures at 200 K.

In WTaV, the strongest negative SRO parameters (representing attraction) are for Ta--V and Ta--W pairs. The varying 1NN and 2NN SRO parameters in the investigated refractory alloys at low temperatures shown in Fig.~\ref{fig:sro} are a strong indication of long-range order~\cite{singh_2015}. Analysing the structure reveals that at low temperatures Ta atoms form almost perfect single \hkl(100) layers, sandwiching two layers of W and V atoms between consecutive Ta layers. The layered structure is evident in the snapshot at 200 K in Fig.~\ref{fig:sro} and is most pronounced at 300 K. Already at 400 K the layering mostly disappears, which is visible as a clear drop in the Ta--Ta SRO parameter in Fig.~\ref{fig:sro}. The binary W--V regions between Ta layers are not ordered with any clear symmetry. The first-nearest neighbour (1NN) SRO W--V parameters are positive but the 2NN parameters are negative.

In WTaCrV, the short-range order produced in the MC/MD simulations is mainly characterised by binary segregation into Cr--Ta regions with almost perfect B2 order and W--V regions with no obvious order (as for WTaV). The order is strong, with significant negative Cr--Ta SRO values up to 1500 K. The SRO parameters are largely consistent with that obtained with another ML potential~\cite{lyu_effects_2023}. However, both our results and those of Ref.~\cite{lyu_effects_2023} are at odds with the experimental results and cluster expansion model in Ref.~\cite{el-atwani_outstanding_2019}, where they instead observed ordering of Cr--V into small precipitates. We will return to this discussion and provide an explanation in Sec.~\ref{sec:prec}.

Short-range order in MoNbTaVW was studied in Ref.~\cite{byggmastar_modeling_2021} with tabGAP modelling but is repeated here for completeness and because the tabGAP used here is an updated (improved) version~\cite{byggmastar_simple_2022} of that in Ref.~\cite{byggmastar_modeling_2021}. The results in Fig.~\ref{fig:sro} are largely consistent with the previous study as well as other studies using cluster expansion Monte Carlo simulations~\cite{toda-caraballo_simulation_2017,fernandez-caballero_short-range_2017}. The SRO in MoNbTaVW mainly consists of Mo--Ta--Nb and W--V regions, with most negative 1NN SRO values for Mo--Ta, W--V, and Mo--Nb pairs, in that order. We also observe that the W--V regions coalesce into roughly $1$ nm thick slabs on a \hkl(111) plane, sandwiching the Mo--Ta--Nb slabs between them. This was first observed by Koskenniemi with the same tabGAP~\cite{koskenniemi_vacancy_2023} but not previously with the older tabGAP version and smaller system size in Ref.~\cite{byggmastar_modeling_2021}. The slab structure is most pronounced at 200 and 300 K and disappears at 400 K (visible as the jump of the V--V SRO value from negative to positive), indicating that the energetic preference of forming slabs with \hkl(111) interfaces is not strong.

\subsection{Segregation}
\subsubsection{Grain boundaries}

\begin{figure}
    \centering
    \includegraphics[width=\linewidth]{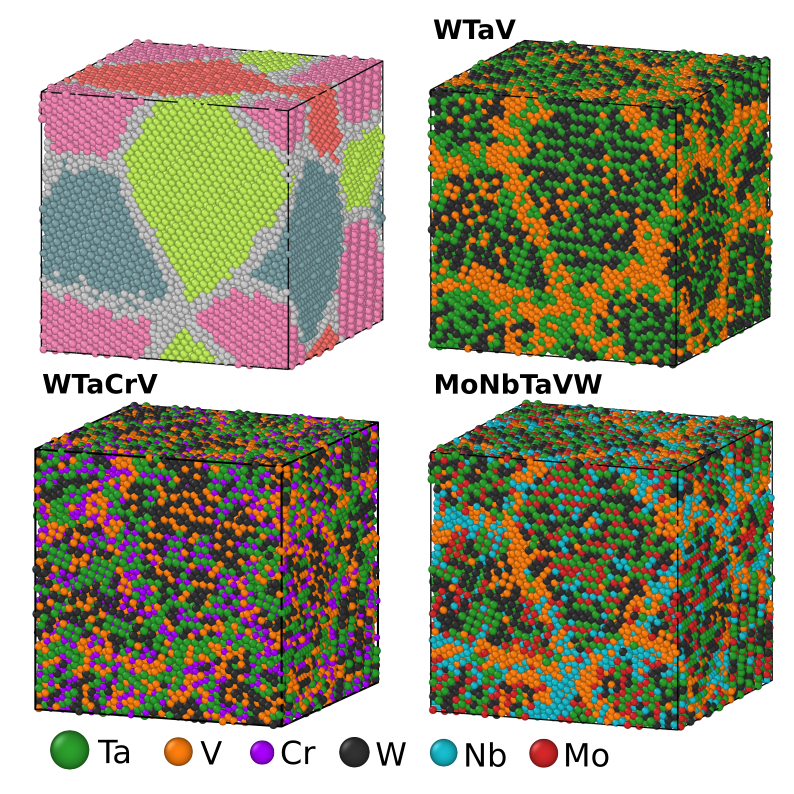}
    \caption{Snapshots of the nanocrystalline alloys, showing the grains and grain boundaries in different colours (top left) and the final segregated WTaV, WTaCrV, and MoNbTaVW systems after MC/MD relaxation at 300 K.}
    \label{fig:poly}
\end{figure}

\begin{table}
    \centering
    \caption{Elemental composition of grain boundary atoms in the three alloys after MC/MD relaxation.}
    \begin{tabular}{lcccccc}
     \toprule
     Alloy & W & Ta & V & Cr & Mo & Nb \\
     \midrule
     WTaV & 0.6 & 32.6 & 66.8 & - & - & - \\
     WTaCrV & 2.1 & 33.2 & 30.7 & 34.1 & - & - \\
     MoNbTaVW & 0.6 & 6.6 & 56.9 & - & 2.1 & 33.9 \\
     \bottomrule
    \end{tabular}
    \label{tab:gbconc}
\end{table}

Figure~\ref{fig:poly} shows snapshots of the nanocrystalline WTaV, WTaCrV, and MoNbTaVW systems after the MC/MD simulation at 300 K. The first snapshot shows the grains and grain boundaries in different colours as identified by the grain segmentation algorithm in \textsc{ovito}. The systems contain four grains in a 10 nm wide cubic box with 61,000 atoms. A larger box (330,000 atoms) was tested and produced qualitatively identical results. Based on the pure-element grain boundary energies in Fig.~\ref{fig:gbs}, it is expected that the energy is minimised by Nb, V, or possibly Ta segregation to the grain boundaries. Indeed, from Fig.~\ref{fig:poly} we see that in WTaV mainly V and some Ta segregate to the grain boundaries. As a more quantitative analysis, Tab.~\ref{tab:gbconc} lists the elemental concentrations of grain boundary atoms. In MoNbTaVW, there is mainly V and Nb segregation. In WTaCrV, there is no obvious segregation, instead only an almost complete depletion of W from the grain boundaries and equal concentrations of the other three elements. Interestingly, the grain boundary composition in WTaV is almost exactly V$_2$Ta, corresponding to the ground-state Laves phases (Fig.~\ref{fig:binaries}). It is possible that grain boundaries can act as nucleation sites for Laves phase growth, even though we did not observe it in the MC/MD simulations.

\subsubsection{Surfaces and voids}

\begin{figure}
    \centering
    \includegraphics[width=\linewidth]{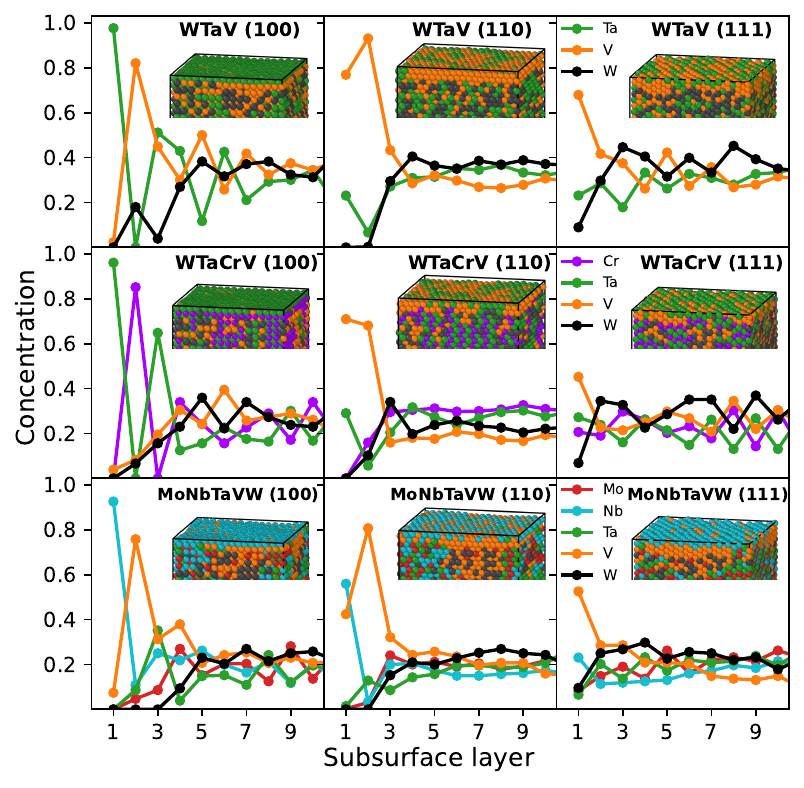}
    \caption{Surface segregation illustrated as layer-by-layer concentration profiles for the three alloys, with snapshots of each surface after MC/MD relaxation shown as insets.}
    \label{fig:surf}
\end{figure}

The size difference and significant variation in surface energies of the elements in the alloys is expected to lead to segregation to surfaces and voids in order to release stress and minimise the energy. Based on Tab.~\ref{tab:props}, pure Nb has the lowest surface energy followed by V and Ta. Fig.~~\ref{fig:surf} shows the results of the MC/MD simulations of three low-index surfaces of the three alloys at 300 K. Indeed, it is mainly Ta, Nb, and V that segregate to the surfaces. In WTaV and WTaCrV, the top \hkl(100) surface layers contain only Ta and for MoNbTaVW only Nb. For the \hkl(110) surface and even more so the \hkl(111) surface, the segregation to the top layer is not as extreme and involve a mix of Ta and V for WTaV and WTaCrV, and Nb and V for MoNbTaVW. The segregation to the subsurface layers appear to be a combined effect of chemical affinity to the topmost-layer elements and minimisation of surface energy. In WTaV, the near-surface layers make up a Ta--V binary region, in WTaCrV mainly Cr--Ta and Ta--V, and in MoNbTaVW Nb--V.

\begin{figure}
    \centering
    \includegraphics[width=\linewidth]{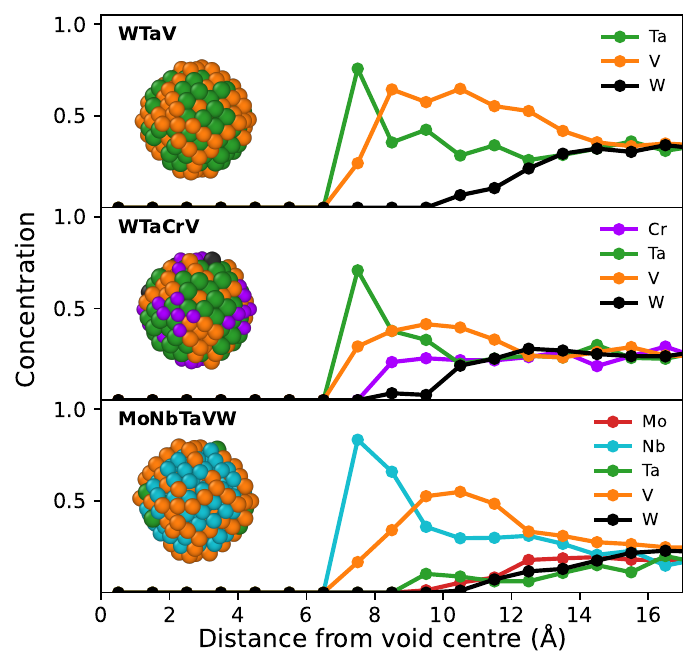}
    \caption{Radial concentrations and snapshots of the elements around voids in the three alloys after MC/MD relaxation.}
    \label{fig:void}
\end{figure}

Fig.~\ref{fig:void} shows the results of segregation around voids inserted in the three alloys and relaxed in MC/MD simulations. Since voids form facets with a mix of different surface orientations inside the crystal, it is natural to expect similar segregation trends as for the flat open surfaces. Fig.~\ref{fig:void} shows that again Ta and V segregate to the inner void surfaces in WTaV, in WTaCrV it is Ta, V and some Cr, and in MoNbTaVW Nb and V (as already observed in Ref.~\cite{byggmastar_modeling_2021}).

\subsubsection{Edge and screw dislocations}

\begin{figure*}
    \centering
    \includegraphics[width=\linewidth]{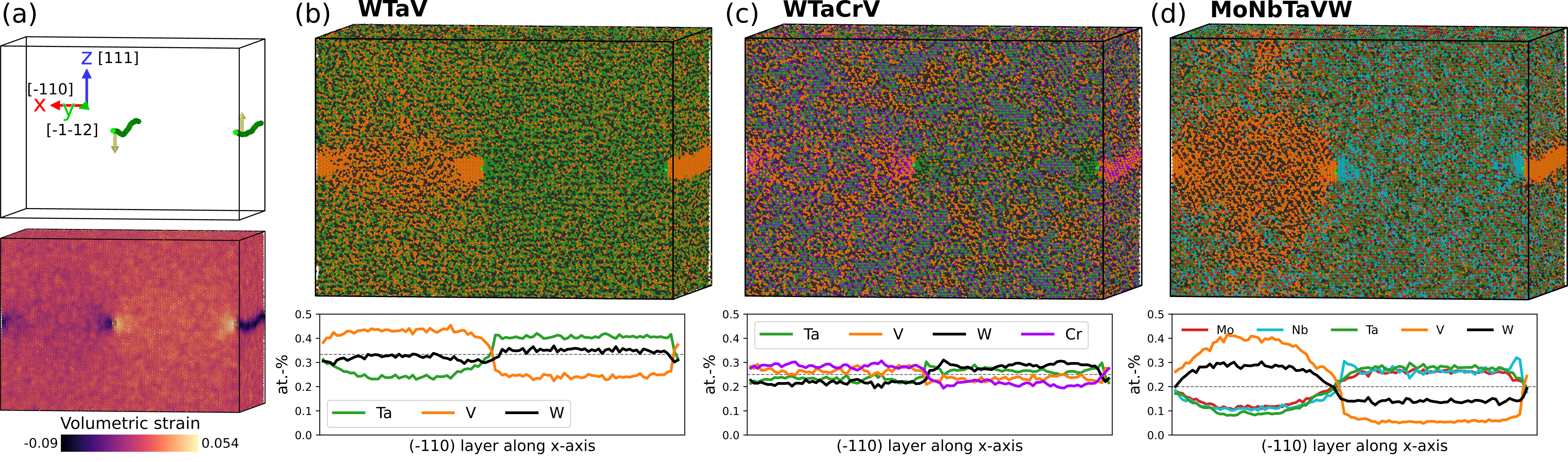}
    \caption{Segregation around edge dislocation lines in the three alloys. (a) shows the edge dislocation dipole structure and the local volumetric strain field in WTaCrV (the other two alloys display similar strain fields). (b), (c), (d) show the final structures after MC/MD simulation. Layer-by-layer atomic composition profiles are shown below each snapshot.}
    \label{fig:edge}
\end{figure*}

Fig.~\ref{fig:edge} shows the results of MC/MD relaxation of 1/2\hkl<111>\hkl(110) edge dislocation dipole structures of the three alloys. We observe that the dislocation cores attract significant segregation, which can be correlated with the strain field induced by the edge dislocations. Fig.~\ref{fig:edge} also shows the volemetric strains around the dislocation lines, showing clear regions of compressive and tensile strain. The snapshots and concentration profiles in Fig.~\ref{fig:edge} show that V, and in WTaCrV some Cr, as the smallest elements segregate to the compressed regions in all three alloys. In contrast, the largest elements Ta (in WTaV and WTaCrV) and Nb in MoNbTaVW segregate to the regions of tensile strain. It is also clear that the segregation is overall much less pronounced in WTaCrV compared to WTaV and MoNbTaVW.

\begin{figure}
    \centering
    \includegraphics[width=\linewidth]{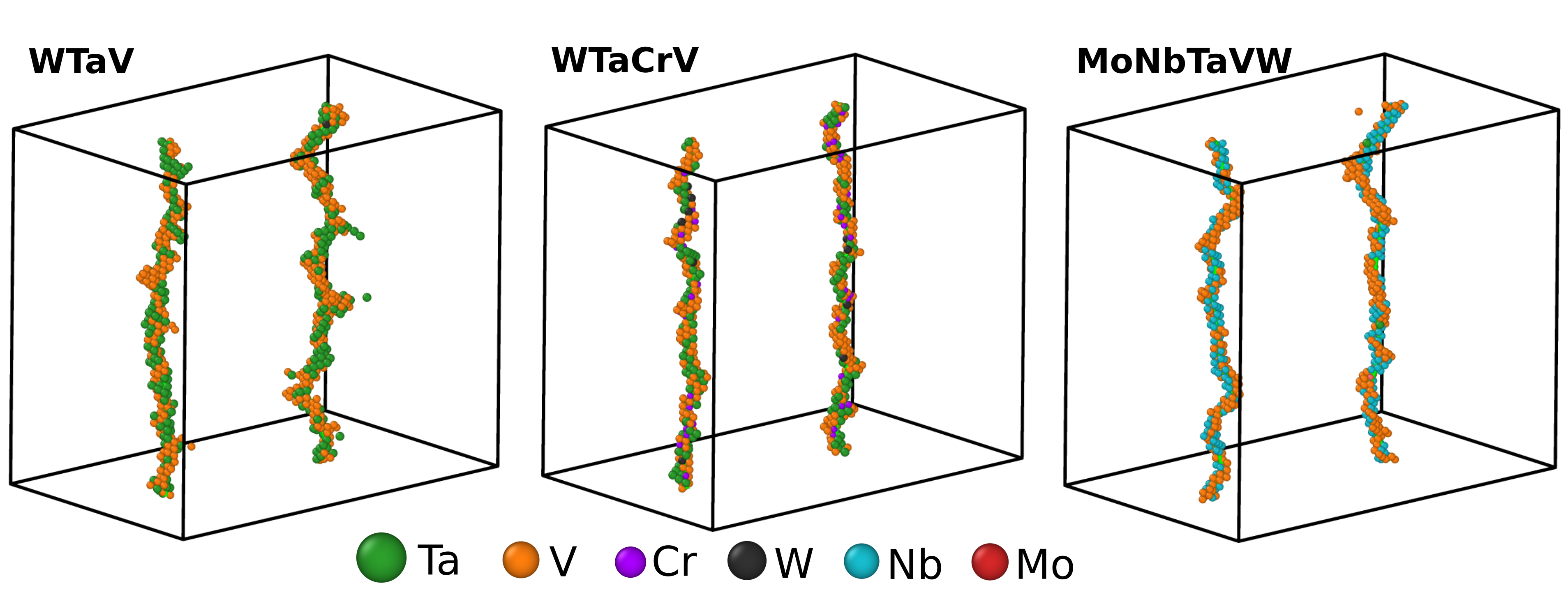}
    \caption{Segregation around screw dislocation cores in the three alloys. Only atoms around the dislocation lines are shown.}
    \label{fig:screw}
\end{figure}

In Fig.~\ref{fig:screw} the results of MC/MD relaxation of screw dislocation dipole structures are shown. The figure shows only the atoms around the dislocation lines. The elemental segregation trends are similar to the edge dislocation case, with the smallest (V) and largest (Ta or Nb) segregating around the screw dislocation core. We also note that the complex chemical surrounding, both before and after segregation, makes the edge and screw dislocation lines wavy or kinked, as is well established~\cite{ma_unusual_2020,yin_ab_2020,maresca_mechanistic_2020,utt_origin_2022,lyu_effects_2023}.

The strong segregation around the dislocation cores, especially in WTaV and MoNbTaVW, also induce fairly long-range chemical ordering effects. In particular, the segregation of V create nucleation spots for extended V--W short-range-ordered regions. This is visible in both the snapshots and the concentration profiles in Fig.~\ref{fig:edge}. Similar extended ordered regions are also present in the screw dislocation structures.

\subsubsection{Dislocation loops}
\label{sec:loop}

\begin{figure}
    \centering
    \includegraphics[width=\linewidth]{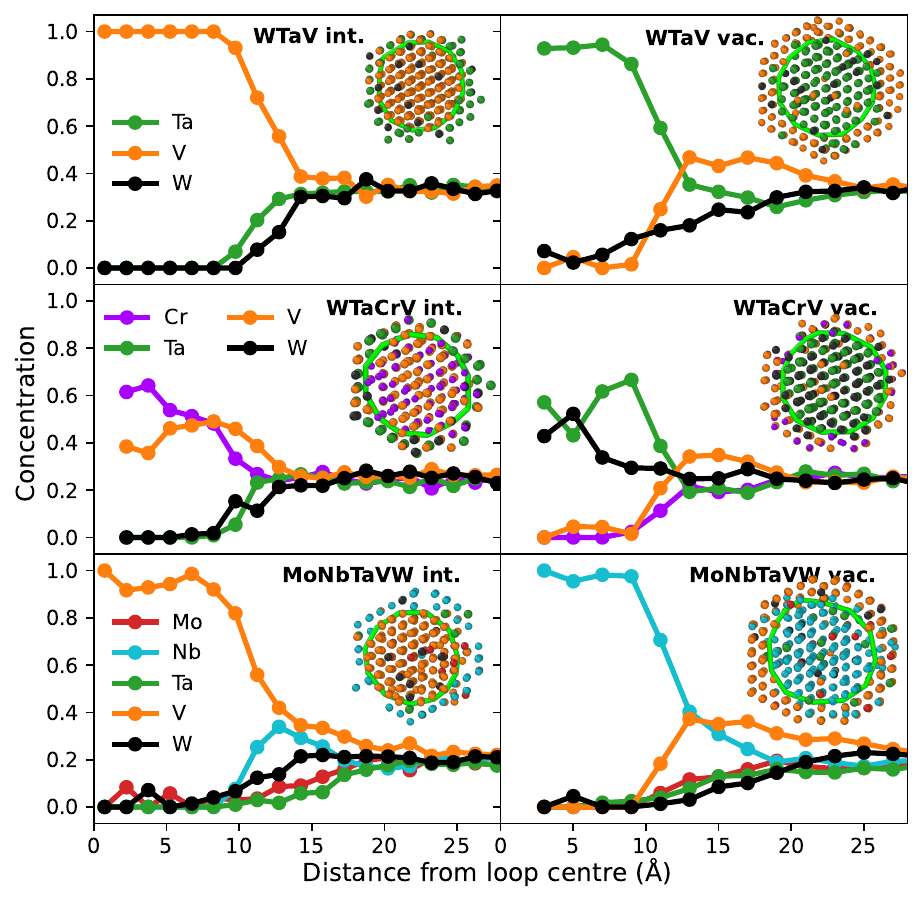}
    \caption{Radial concentration profiles showing the segregation in and around interstitial and vacancy 1/2\hkl<111> dislocation loops. The insets show snapshots of the dislocation loops with surrounding atoms.}
    \label{fig:loop}
\end{figure}

Fig.~\ref{fig:loop} shows the results from MC/MD relaxation of both interstitial- and vacancy-type 1/2\hkl<111> prismatic dislocation loops. The results are summarised in plots of the radial concentration profiles from the centre of the loop. The same trends as for infinite edge dislocation lines are apparent: V (and Cr in WTaCrV) segregate to the compressed regions. For interstitial-type loops this is the lattice cylinder encircled by the dislocation line and for vacancy-type loops it is the periphery of the dislocation line. Hence, the segregation trends of interstitial and vacancy loops are essentially each others inverse, following the geometry and strain fields of dislocation loops. It is again worth noting that while in WTaV and MoNbTaVW the segregation is elemental, i.e. pure V or Ta (in WTaV) and Nb (in MoNbTaVW), in WTaCrV the segregation is more complex and involves two elements each, V and Cr to compressed regions and Ta and W to spacious regions.

\subsection{Interfaces and precipitates}
\label{sec:prec}

\subsubsection{Atom probe tomography analysis}

There is a remarkable discrepancy between the predicted short-range order in single crystal WTaCrV by ML potentials (both here in Fig.~\ref{fig:sro} and in Ref.~\cite{lyu_effects_2023}) and that produced by the CE model in Ref.~\cite{el-atwani_outstanding_2019}. The latter predicts formation of almost pure binary clusters of Cr ($\sim$65\%) and V ($\sim$30\%) in a surrounding WTa-rich matrix, which agrees with the experimental observation of Cr-rich precipitates in WTaCrV~\cite{el-atwani_outstanding_2019}. In complete contrast, the ML potential here and in Ref.~\cite{lyu_effects_2023} produce CrTa and WV binary regions. Thermodynamically this is not expected since Fig.~\ref{fig:binaries} shows that the combination of CrV and WTa binaries has lower formation energy ($-0.08$ eV/atom) than combining CrTa and WV ($-0.05$ eV/atom). However, interfacial energies and strain energies due to lattice mismatch may change the preferential interfaces at short length scales, which requires detailed analysis.


\begin{figure*}
    \centering
    \includegraphics[width=\linewidth]{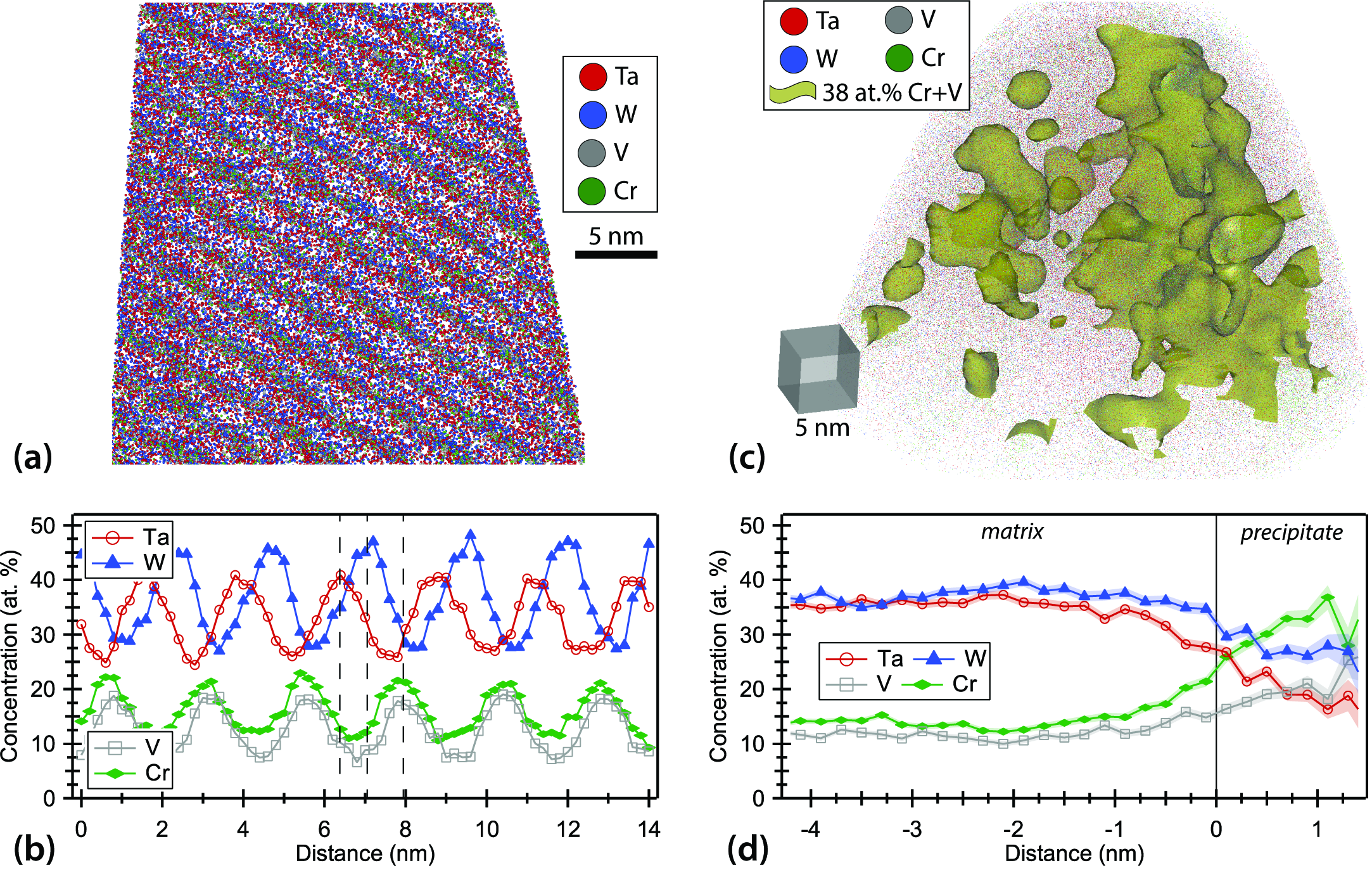}
    \caption{APT analyses of (a, b) as-deposited W$_{38}$Ta$_{36}$Cr$_{15}$V$_{11}$ material and (c, d) the same material after irradiation. Both the atom map (a) and 1D concentration profile (b) of the as-deposited material reveal a striated composition, with peak concentrations of Ta and W offset from one another and from the V and Cr profile. (c) Isoconcentration surfaces reveal globular precipitates enriched in Cr and V after irradiation, as further quantified by a proximity histogram in (d). Image depth of (a) is 5 nm. Error in (d) represents 1$\sigma$ from standard counting error. Error bars in (b) are smaller than the symbol size.}
    \label{fig:expt}
\end{figure*}

Before theoretically analysing the interface energetics, we obtain further information about the length scales and concentration profiles of the interface- and precipitate-rich thin films from Ref.~\cite{el-atwani_outstanding_2019} to allow closer comparison between predictions and experiments. We used atom probe tomography (APT) to analyse the same W$_{38}$Ta$_{36}$Cr$_{15}$V$_{11}$ thin film samples as in Ref.~\cite{el-atwani_outstanding_2019}. APT was used to determine the local 3D elemental distribution in both the as-deposited and irradiated materials. Representative reconstructions and associated local composition profiles are provided in Fig.~\ref{fig:expt}. For the as-deposited material, the APT atom map reconstruction (Fig.~\ref{fig:expt}a) reveals laminar nanoscale compositional modulations in the Ta, W, V and Cr distribution. These compositional variations were quantified using a 1D concentration profile along a 23 nm diameter cylindrical volume in Fig.~\ref{fig:expt}b. In this profile, Cr and V are co-segregated and vary in local composition by $\pm 5$ at.\%. Stronger variations are found for W ($\pm 10$ at.\%) and Ta ($\pm 8$ at.\%), which are also slightly offset from one another. The approximate peak concentrations are noted by vertical dashed lines in Fig.~\ref{fig:expt}b to visually identify this offset of each elemental profile. The APT reconstruction and associated isoconcentration surfaces of Cr+V in Fig.~\ref{fig:expt}c for the same material after irradiation reveals that the striated compositional variations have been replaced by more conventional globular precipitates. A proximity histogram in Fig.~\ref{fig:expt}d reveals local enrichment of Cr (up to 35 at.\%) and V ($\sim$22 at.\%), which is significantly stronger than the laminar compositional variations found in the as-deposited material.

\subsubsection{Coherent and semicoherent interfaces}

\begin{figure*}
    \centering
    \includegraphics[width=0.7\linewidth]{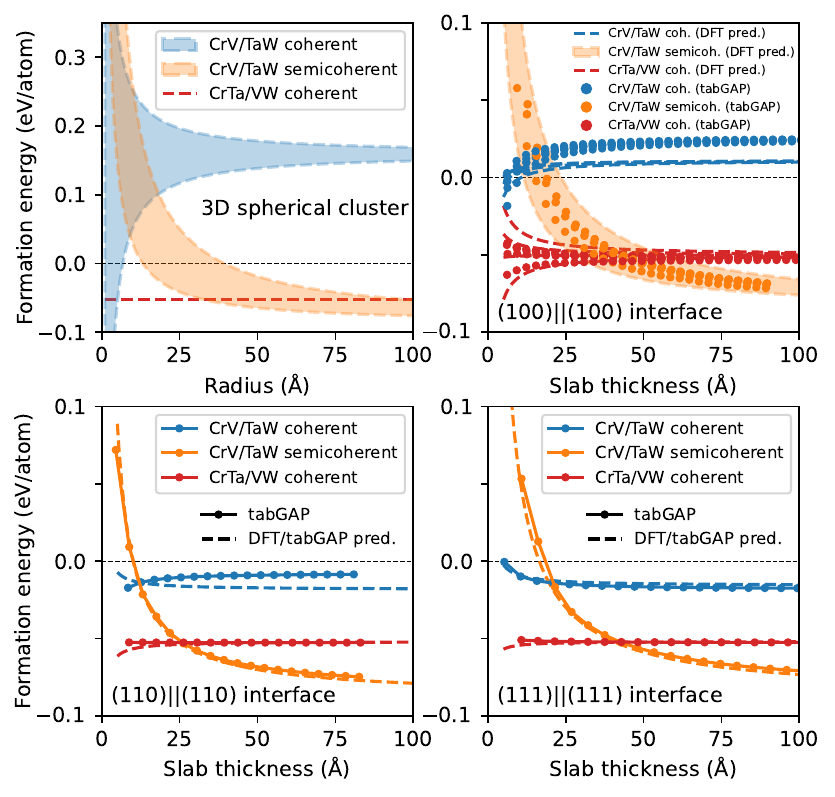}
    \caption{Formation energies per atom of coherent and semicoherent 3D and 2D periodic interfacial structures of B2 binaries as functions of radius or slab thickness. The data points and solid lines are results from direct relaxations with the tabGAP. Dashed lines are DFT/tabGAP-based predictions based on Eq.~\ref{eq:Ef_int} using DFT data from Tab.~\ref{tab:coh} and tabGAP semicoherent interfacial energies. The shaded regions span predictions using the lowest and highest interfacial energies. The different lines and data points for the \hkl(100) interface are the four different interface chemical terminations from Tab.~\ref{tab:coh}.}
    \label{fig:intf}
\end{figure*}

Fig.~\ref{fig:expt} shows that both the CrV-rich precipitates in the irradiated sample and the CrV- and TaW-rich layers in the as-deposited sample are only 2--5 nm in length. To analyse the interface energetics, we choose as a model system equiatomic CrV/TaW interfaces with both phases in B2 order. This is the closest system to the experimental structure that is well-defined enough for DFT calculations with small supercells. From Fig.~\ref{fig:B2}, the equilibrium atomic volume of CrV is 12.2 Å$^3$ and for TaW 17 Å$^3$, which is a significant enough mismatch that semicoherent interfaces with misfit dislocations must be considered. We also consider the competing thermodynamically stable CrTa/VW interface system, where the lattice mismatch is negligible and only coherent interfaces are expected. We analyse three different low-index interfaces, $\hkl(100) \parallel \hkl(100)$, $\hkl(110) \parallel \hkl(110)$, and $\hkl(111) \parallel \hkl(111)$ as well as 3D spherical coherent and semicoherent clusters or precipitates. The objective of our analysis is two-fold: (1) explain why the ML potential predicts CrTa/VW clustering instead of the experimentally observed CrV/TaW, and (2) quantify the competition between coherent and semicoherent interfaces and precipitates as a function of length for comparison with the APT results.

\begin{table}
    \centering
    \caption{Properties of coherent interfaces. Coherency strain energies, $E_\mathrm{CS}$ (eV/atom), for three different interface orientations and the coherent shared lattice constant perpendicular to the interface, $a_\perp$ (Å). The same properties are also given for the 3D case, i.e. isotropically strained cluster for coherency in all directions. $\sigma$ (eV/Å$^2$) are interfacial energies. All properties are computed with the tabGAP and in DFT. For $\hkl(100) \parallel \hkl(100)$ interfaces, four different values are given for the four different possible interface configurations between the two binaries. For CrV/TaW interfaces, the \hkl(100) atomic layer orders are VCr$|$WTa, CrV$|$TaW, VCr$|$TaW, and CrV$|$WTa for the four values, respectively, where $|$ marks the interface. For CrTa, they are VW$|$CrTa, WV$|$TaCr, WV$|$CrTa, and VW$|$TaCr.}
    \begin{tabular}{lrrrr}
     \toprule
     & \multicolumn{2}{c}{CrV/TaW} & \multicolumn{2}{c}{CrTa/VW} \\
     \cmidrule{2-3} \cmidrule(lr){4-5}
     & tabGAP & DFT & tabGAP & DFT \\
     \midrule
     $\Delta E_\mathrm{CS}^{\hkl(100)}$ & 0.111 & 0.097 & 0 & 0 \\
     $\Delta E_\mathrm{CS}^{\hkl(110)}$ & 0.078 & 0.069 & 0 & 0 \\
     $\Delta E_\mathrm{CS}^{\hkl(111)}$ & 0.067 & 0.072 & 0 & 0 \\
     $a_\perp$ & 3.13 & 3.13 & 3.09 & 3.09 \\
     $\Delta E_\mathrm{CS}^\mathrm{3D}$ & 0.249 & 0.248 & 0 & 0 \\
     $a_\perp^\mathrm{3D}$ & 3.14 & 3.13 & 3.09 & 3.09 \\
     $\sigma_{\hkl(100)}$ & $-0.055$ & $-0.047$ & $-0.044$ & $-0.050$ \\  
      & 0.030 & 0.039 & 0.051 & 0.049 \\  
      & 0.005 & 0.013 & 0.174 & 0.172 \\  
      & $-0.033$ & $-0.023$ & $-0.157$ & $-0.158$ \\  
     $\sigma_{\hkl(110)}$ & $-0.005$ & 0.004 & 0.000 & $-0.003$ \\
     $\sigma_{\hkl(111)}$ & 0.006 & 0.004 & 0.002 & $-0.002$ \\
     \bottomrule
    \end{tabular}
    \label{tab:coh}
\end{table}

Table~\ref{tab:coh} shows the coherency strain energies, coherent lattice constants, and coherent interfacial energies calculated with both the tabGAP and DFT from Eqs.~\ref{eq:ef} and \ref{eq:Ef_int}. Overall, the tabGAP results agree well with the DFT calculations. For $\hkl(100) \parallel \hkl(100)$ interfaces we considered the different chemical terminations at the interface layer. For \hkl(110) and \hkl(111) we considered two different interface chemical orders, but both yielded interfacial energies very close to zero so we only provide one value in Tab.~\ref{tab:coh}. Semicoherent interfacial energies, which include the energy for accommodating misfit dislocations and the actual interface, for CrV/TaW are only accessible with the tabGAP. We get 0.05-0.15 eV/Å$^2$ for \hkl(100) with the different interface terminations, 0.057 eV/Å$^2$ for \hkl(110), and 0.096 eV/Å$^2$ for \hkl(111). That \hkl(100) and \hkl(110) interfacial energies are lowest in energy is consistent with recent experiments that observed these two interfaces between a B2 and BCC phase in another refractory alloy~\cite{kloenne_bcc_2023}.

The data in Tab.~\ref{tab:coh} can be used in Eq.~\ref{eq:Ef_int} to quantify the competition in formation energy between the different interfaces as a function of interface slab thickness. With the tabGAP, the formation energies can be computed directly by relaxing interfacial structures of different size. Fig.~\ref{fig:intf} shows the results for the three considered interfaces as well as for the 3D case of spherical clusters. For the 3D case we use Eq.~\ref{eq:Ef_int} with the coherency strain energies and the coherent interfacial energies from DFT. Semicoherent interfacial energies are always from tabGAP (hence the label ``DFT/tabGAP pred.'' in Fig.~\ref{fig:intf}). The minimum and maximum interfacial energies are used to show a range of uncertainty. The limiting values at large radii or slab thickness in Fig.~\ref{fig:intf} are from Eq.~\ref{eq:Ef_int} $\zeta + \Delta E_\mathrm{f}^{AB} / N$, i.e. the strain energy plus the formation energies of the two B2 binaries. $\zeta = 0$ for semicoherent interfaces and $\zeta = \Delta E_\mathrm{CS}$ for coherent interfaces. $\Delta E_\mathrm{f}^{AB} / N$ is the formation energy of the two B2 slabs in their bulk relaxed state, which are $-0.08$ eV/atom for CrV combined with TaW in both DFT and tabGAP and $-0.05$ eV/atom for CrTa and WV. Hence, it is clear that semicoherent CrV/TaW configurations will eventually be thermodynamically favoured at sizes where interfacial area becomes negligible compared to the system volume.

Fig.~\ref{fig:intf} explains why CrTa/VW clustering is observed in MC/MD simulations of a coherent WTaCrV lattice. For both 3D and 2D interfacial configurations, coherent CrTa/VW systems are much lower in formation energy than coherent CrV/TaW. For thick interface slabs or large precipitates, semicoherent CrV/TaW configurations become most favourable. The crossover between semicoherent and coherent CrV/TaW interfaces occurs at a slab thickness of 1--2 nm and between semicoherent CrV/TaW and coherent CrTa/VW at 2.5--3 nm. For 3D clusters, the uncertainty is large but the lowest crossover between semicoherent CrV/TaW and coherent CrTa/VW occurs at 3 nm in radius. Coherent CrV 3D precipitates in TaW are thermodynamically unstable at all sizes. Hence, Fig.~\ref{fig:intf} suggests that the CrV-rich precipitates visible in the APT analysis in Fig.~\ref{fig:expt} must be semicoherent. Regarding the laminar interfacial structure of the as-deposited sample in Fig.~\ref{fig:expt}, the layer thickness is similar to the crossovers between coherent and semicoherent interface slabs in Fig.~\ref{fig:intf}, making it difficult to draw definitive conclusions. Additionally, the theoretical crossovers are for ideal equiatomic B2 binaries, while the concentration profiles in Fig.~\ref{fig:expt} show more chemical variations. Nevertheless, since only CrV/TaW slabs and precipitates are observed and not CrTa/VW, Fig.~\ref{fig:intf} strongly suggests that also the pre-irradiated interfaces are semicoherent.

\begin{figure*}
    \centering
    \includegraphics[width=\linewidth]{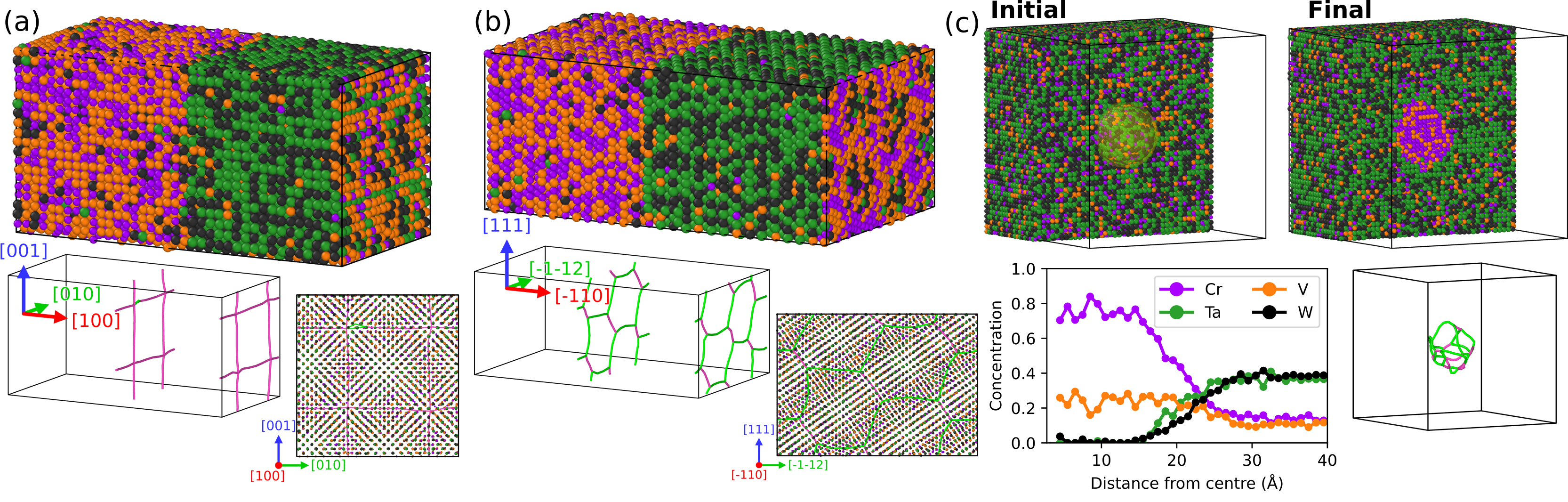}
    \caption{Semicoherent interfaces between CrV and WTa after MC/MD relaxation, (a): \hkl(100)/\hkl(100) and (b): \hkl(110)/\hkl(110). The smaller snapshots show side views and cross sections of the misfit dislocation grids with pink lines representing \hkl<100> Burgers vectors and green lines 1/2\hkl<111>. The spacing between misfit dislocation lines in the interface plane is about 3 nm. (c) Semicoherent precipitate in WTaCrV before and after MC/MD relaxation, shown as half-box slices and the misfit dislocation network after relaxation. The plot shows the radial concentration profile starting from the centre of the relaxed precipitate.}
    \label{fig:prec}
\end{figure*}

To get more insight into the chemical order of interfaces, we create \hkl(110) and \hkl(100) interfaces with slab widths of 6 nm, both coherent and semicoherent, and let them evolve in MC/MD simulations. The temperature in both MC and MD is 1000 K, close to the experiments~\cite{el-atwani_outstanding_2019}. The coherent interfaces quickly transform from CrV/WTa to CrTa/WV-ordered mixtures after MC swaps, as is expected from the single-crystal simulations in Sec.~\ref{sec:sro} and Fig.~\ref{fig:intf}. For the semicoherent CrV/WTa interfaces, both regions remain as binary CrV and WTa regions, with only some impurities and the initial B2 order changing to a more random order due to the high temperature. Figures~\ref{fig:prec}(a)-(b) show the final semicoherent interfaces after MC/MD relaxation. 

As a simulation more representative of the real CrV-rich precipitates in WTaCrV, we also embedded an initially incoherent spherical BCC precipitate in a surrounding BCC lattice. The diameter of the precipitate is 4 nm, similar to the observed precipitates~\cite{el-atwani_outstanding_2019}. The lattice constant of the precipitate is smaller (2.6 Å) than the surrounding matrix (3.2 Å). Instead of starting from a CrV precipitate, we started with the entire system in random solution with the same composition as the experiments, W$_{38}$Ta$_{36}$Cr$_{15}$V$_{11}$~\cite{el-atwani_outstanding_2019}. Following MC/MD evolution, Cr and V quickly cluster to the compressed precipitate, creating a semicoherent precipitate as shown in Fig.~\ref{fig:prec}(c). This is in perfect agreement with the experiments and CE model~\cite{el-atwani_outstanding_2019} and expected based on Fig.~\ref{fig:intf}. Figure~\ref{fig:prec}(c) also shows a radial concentration profile of the precipitate. The atomic composition of the precipitate after MC/MD relaxation is approximately 70\% Cr, 30\% V. This is also in quantitative agreement with Ref.~\cite{el-atwani_outstanding_2019}. The interface can be characterised by a network of misfit dislocations, most of which are 1/2\hkl<111> as visualised in Fig.~\ref{fig:prec}(c).

From this we conclude that CrV-rich precipitates in WTaCrV must be semicoherent. Finally, it is worth noting why the CE modelling in Ref.~\cite{el-atwani_outstanding_2019} successfully reproduced the experimental observations despite modelling a coherent BCC lattice. The CE model is fitted to a database of relaxed structures of WTaCrV alloys. Even though the simulations assume a rigid coherent lattice, the model will reproduce the thermodynamically stable short-range order. This means that the energetically favoured CrV and WTa clustering evolves because the CE model implicitly accounts for local relaxation, even though there is no explicit relaxation like in MD. As such, the results represent relaxed CrV precipitates, neglecting the interface structure whether it is coherent or not. In MD, the single-crystal simulation is constraint in a single coherent lattice with no energetically available pathway to form a semicoherent CrV precipitate, and only the analysis above and starting simulations from in- or semi-coherent interfaces allow us to correctly reproduce the experimental observations.

\section{Discussion}

The striated structure of the as-deposited W$_{38}$Ta$_{36}$Cr$_{15}$V$_{11}$ thin film revealed by the APT analysis in Fig.~\ref{fig:expt} is qualitatively similar to the interface models prepared and equilibrated in the simulations in Fig.~\ref{fig:prec}. Our calculations predict that the most stable ideal CrV/WTa interface is semicoherent with \hkl(110)/\hkl(110) orientations when the interfacial slabs are thicker than 2--3 nm. For coherent interfaces, CrTa/VW-like structures are more stable and emerge in MC/MD simulations of a coherent lattice, contradicting the experimental observations. Our analysis in Fig.~\ref{fig:intf} thus highlights the importance of considering both coherency strain and strain-free thermodynamical stability when understanding and predicting short-range order and interface structures. Irradiation of W$_{38}$Ta$_{36}$Cr$_{15}$V$_{11}$ resulted in more globular but small CrV-rich precipitates (Fig.~\ref{fig:expt}), which our calculations and simulations predict to be semicoherently embedded in the surrounding matrix. Even though no direct high-resolution imaging of the interface structures was possible to confirm this, the concentration profiles from the APT analysis are consistent with the simulated thermodynamically equilibrated precipitates. Both simulation and experiment show that the chemical composition of the precipitates is mainly CrV, with more Cr than V.

In arc-melting of bulk high-entropy alloy samples, the vastly different melting points of refractory metals often leads to elemental segregation during the solidification. In MoNbTaVW, a dendritic structure is easily formed with segregation of Mo, Nb, and V to the dendrite boundaries and W to dendrite cores, with Ta uniformly distributed~\cite{senkov_refractory_2010}. This was presumed to be due to the difference in melting points~\cite{senkov_refractory_2010}. Our results show that there is also a strong thermodynamic driving force for segregation of Nb and V to grain boundaries, since they not only have the lowest melting points but also the lowest grain boundary energies. Segregation of Nb to grain boundaries was also previously observed in simulations of the similar alloy MoNbTaW~\cite{li_complex_2020,aksoy_chemical_2022}. In the WTaCrV thin films produced by magnetron sputtering, evidence of some segregation of Cr, V and some Ta to grain boundaries was found~\cite{el-atwani_outstanding_2019}. This is consistent with our results of depletion of W from grain boundaries in the MC/MD simulations, which leaves Cr, V, and Ta at the boundaries.

Overall, the energetically favoured segregation to the various defects in the three alloys investigated here can be linked to fundamental material properties. Grain boundary segregation almost directly correlates with pure-element grain boundary energies. Surface and void segregation correlates with pure-element surface energies. Segregation to edge and screw dislocation lines and loops is mainly driven by minimising the strain, so that small atoms (Cr, V) prefer regions of compressive strain and large atoms (Ta, Nb) regions of tensile strain. These observations make it possible to, with reasonable confidence, predict the energetically preferred segregation trends also in other refractory alloys. Nevertheless, it is not entirely that trivial. Thermodynamic affinity between elements and strain can also complicate the segregation preference. For example, Cr remains at grain boundaries in WTaCrV despite having similar grain boundary energies to W and twice as high as V and Ta. This is driven by the stability of CrV mixtures and presumably also the small size of Cr. The correlation between surface segregation and element with lowest surface energy in pure metal form is strong, yet not perfect. Different surface geometries, similar surface energies of some elements, and thermodynamics affect the preferential surface segregation. In WTaV and WTaCrV, the top surface layer for the \hkl(100) is pure Ta, but for \hkl(110) and \hkl(111) the top layer is mainly V.

We observe that segregation can also act as a nucleation and centre of enhanced short-range order regions. This is particularly clear for edge dislocations in Fig.~\ref{fig:edge}. In WTaV and MoNbTaVW, the segregation of V to the strained lattice beside the dislocation line attracts W to form large regions of WV short-range order that extend far from the dislocation line. More generally, this suggests that immobile defects may attract not only elemental segregation but also be further surrounded by strongly short-range-ordered regions. The consequences of this, for example for the mobility of dislocations upon applied stress or increased temperatures may be significant and requires further investigation.

The same applies to our observation in Sec.~\ref{sec:prec} that the CrV--rich precipitates in WTaCrV observed in Ref.~\cite{el-atwani_outstanding_2019} might not be coherent but form semicoherent interfaces with the surrounding matrix. How this affects the mechanical properties of the WTaCrV alloys is unclear. What also remains unclear is the formation mechanism of the semicoherent CrV precipitates. Our results demonstrate that they are thermodynamically very stable, but the kinetics of formation is difficult to model on the scale of MD simulations. Other, more recent experimental studies of WTaCrV alloys did not observe any precipitates or secondary phases~\cite{shi_deuterium_2022,shi_helium_2023,kalita_microstructure_2023}, suggesting that the formation may be sensitive to synthesis method and conditions, and possibly alloy composition. One possible formation mechanism based on the results in Sec.~\ref{sec:loop} is that they nucleate from interstitial clusters. We found that Cr and V segregate to interstitial dislocation loops and there are geometric similarities between a prismatic interstitial dislocation loop and a semicoherent precipitate. Single interstitials are also preferentially Cr- or V-containing dumbbells (Fig.~\ref{fig:vacmig}), supplying a pathway for growth and nucleation of CrV precipitates by interstitial migration.

\section{Conclusions}

We have thoroughly investigated short-range order and segregation in the three refractory alloys WTaV, WTaCrV, and MoNbTaVW in hybrid MC/MD simulations. To this end, we developed a new machine-learned interatomic potential for the full composition space of the W--Ta--Cr--V system and showed that it accurately reproduces properties relevant for ordering and segregation in WTaCrV alloys. Even though we here used to the potential to study mainly equiatomic refractory alloys, the potential is trained to also be applicable to alloy composition far from equiatomic compositions. Importantly, the potential also correctly reproduces the high stability of intermetallic Laves phases of TaV$_2$ and TaCr$_2$. We found that in nanocrystalline WTaV, segregation leads to local concentrations close to TaV$_2$ at grain boundaries, which could trigger Laves phase nucleation and growth.

Our main conclusion is that although many segregation trends can be trivially linked to fundamental properties (such as grain boundary energies, surface energies, or atom size), not all segregation and ordering results can, and depend on additional thermodynamic properties of the alloy system such as favourable or unfavourable mixing of certain elements. Among the three studied alloys, we found that WTaCrV is most resistant to elemental segregation due to competing thermodynamic mixing. For example, Cr segregation to interstitial dislocation loops or precipitates is accompanied by V to form stable CrV mixtures, while in MoNbTaVW and WTaV segregation is often elemental (e.g. pure V to compressed lattice regions and pure Ta or Nb to open-volume defects). Our results therefore show that basic material properties can be used to predict some general segregation trends in arbitrary refractory alloy compositions, but a detailed quantitative prediction requires full large-scale atomistic simulations. Our results also highlight the intricate competition between coherent and semicoherent interfaces in WTaCrV and we  predict that the CrV-rich precipitates observed experimentally are semicoherently embedded in the TaW-rich matrix.

\section*{Acknowledgements}

Arun Devaraj (PNNL) is acknowledged for preparing and collecting initial APT data.
JB was supported by funding from the Research council of Finland through the OCRAMLIP project, grant number 354234.
Grants of computer capacity from CSC - IT Center for Science are gratefully acknowledged.
JSW and DS work was supported by the National Science Centre, Poland, under research project no UMO-2019/35/D/ST5/03526.
DNM work has been carried out within the framework
of the EUROfusion Consortium and has received
funding from the Euratom research and training programme
2014-2018 and 2019-2020 under grant agreement
No 633053. The views and opinions expressed herein
do not necessarily reflect those of the European Commission.
DNM also received funding from the Euratom
research and training programme 2019-2020 under
grant agreement No. 755039. We acknowledge funding
by the EPSRC Energy Programme [grant number
EP/W006839/1].
OEA, DKS, and EM  acknowledge funding from the U.S. Department of Energy, Advanced Research Projects Agency-Energy program under contract number DE-AR0001541.

\appendix
\section{Effect of magnetism}
\label{sec:app-mag}

\begin{figure}
    \centering
    \includegraphics[width=\linewidth]{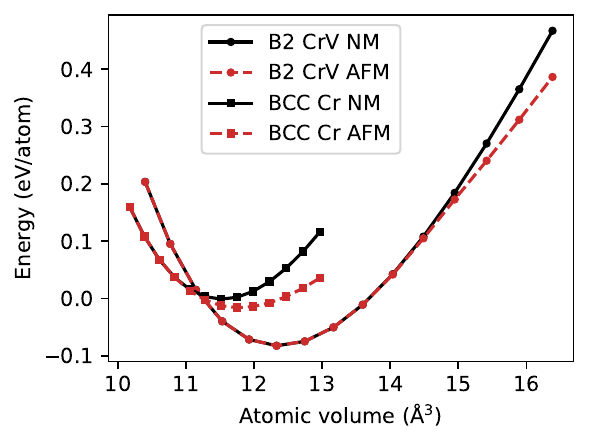}
    \caption{Energy versus volume for pure Cr and B2 CrV alloys with (AFM) and without (NM) spin-polarised DFT calculations. The energies are shifted by the energies of relaxed pure Cr and V from non-spin-polarised calculations.}
    \label{fig:mag}
\end{figure}

We here analyse the effects of neglecting magnetism in the ML potential and its DFT training data. Figure~\ref{fig:mag} shows the energy-volume curves for pure Cr and B2 CrV from non-spin-polarised DFT (NM) and from calculations with initially antiferromagnetic states (AFM). The energies are shifted with respect to nonmagnetic relaxed pure element energies. Figure~\ref{fig:mag} shows that magnetic Cr is elastically noticeably softer when stretched even though the relaxed atomic volume and energies are close. The equilibrium AFM energy of Cr is only 0.015 eV/atom lower than NM Cr, which is similar to previous results~\cite{hafner_magnetic_2002,pankhurst_electronic_2004}. For compressed lattices, the magnetic order vanishes. The ML potential here trained to nonmagnetic Cr will thus show largest errors with respect to the magnetic ground state for stretched pure Cr or dilute Cr alloys. However, Fig.~\ref{fig:mag} also shows that already for the binary B2-ordered CrV alloy, the magnetic moments vanish except for very large atomic volumes. We also confirmed that this is true for the second-most stable CrV phase. Since large CrV atomic volumes are still expected in WTaCrV alloys and for coherent CrV/TaW interfaces, we also computed the coherency strain energies $\Delta E_\mathrm{CS}$ for the CrV/TaW \hkl(100) interface and 3D coherent cluster. For the \hkl(100) interface, the initialised AFM moments vanished during relaxation and $\Delta E_\mathrm{CS}$ is identical to the value in Tab.~\ref{tab:coh}. For the 3D case, $\Delta E_\mathrm{CS}$ decreases from 0.248 eV/atom to 0.230 eV/atom with spin-polarised calculations. Using this value in Fig.~\ref{fig:intf} leads to only a slight decrease in the 3D CrV/TaW coherent cluster formation energy and all conclusions remain the same.

\section{Cross-validation of ML potentials}
\label{sec:app-cross}

\begin{figure}
    \centering
    \includegraphics[width=\linewidth]{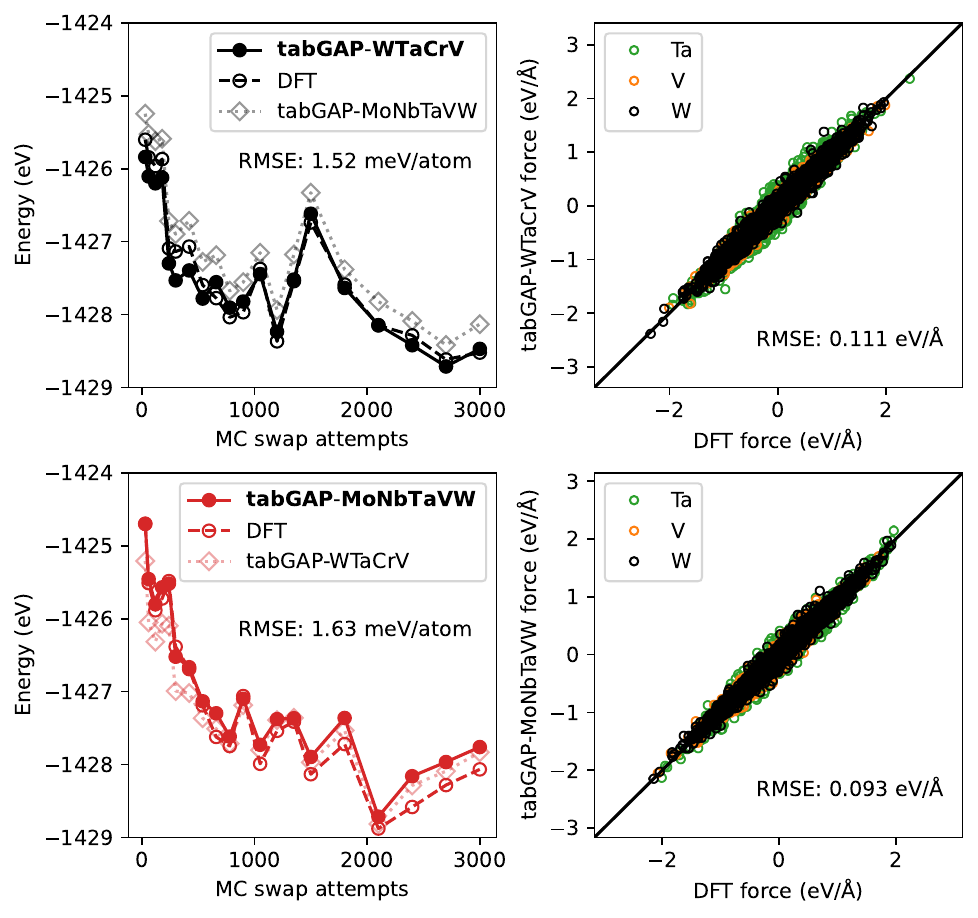}
    \caption{Cross-validation of the new W--Ta--Cr--V tabGAP and the previous Mo--Nb--Ta--V--W tabGAP for WTaV alloys and comparison with DFT. The left plots show MC/MD simulations of a WTaV system with one vacancy with each tabGAP. The energies and forces are recomputed with the other tabGAP and DFT and shown as comparison. The tabGAP indicated in bold font is the one used in the MC/MD simulation and in the RMSE calculation and comparison to DFT. The right plots shows the force component comparison between the tabGAP and DFT for all frames.}
    \label{fig:vasp_WTaV}
\end{figure}

Both the tabGAP developed here for W--Ta--Cr--V alloys and the previously developed Mo--Nb--Ta--V--W tabGAP~\cite{byggmastar_simple_2022} can be used to simulate WTaV alloys. The two potentials are similar, as we used the same hyperparameters and shared training data, although the new potential includes more data for pure metals and ordered binaries. Here we cross-validate the two tabGAPs for WTaV between each other and DFT in MC/MD simulations. As a simple test case that includes both short-range order and some segregation, we perform MC/MD simulations of an initially randomly ordered WTaV alloy containing one vacancy in a $4\times4\times4$ lattice (127 atoms). MC/MD simulations are done with both tabGAPs at 300 K. Then, we extracted frames from both simulations at regular intervals. Energies and forces of those frames were calculated statically with DFT and the tabGAP that was not used in the MC/MD simulation. Finally the two tabGAPs and DFT can be validated against each other. The results are shown in Fig.~\ref{fig:vasp_WTaV}, where the energy and force RMSEs are shown for both tabGAPs against DFT for all frames. The results show that the accuracy for both tabGAPs is the same, 1.5--1.6 meV/atom for energy and 0.1 eV/Å for force. The energy as functions of MC step also show similar trends in both tabGAPs and DFT. From this we conclude that both tabGAPs are (1) accurate with respect to DFT for ordering and segregation in WTaV, and (2) similar, so that both tabGAPs are expected to produce the same segregation and ordering results for WTaV and other shared alloy compositions.

%

\end{document}